\documentclass[a4paper,11pt]{article}
\usepackage{jheppub} 
\pdfoutput=1
\usepackage{graphicx,array}
\usepackage[dvipsnames]{xcolor}
\usepackage{amstext}
\usepackage{amssymb}
\usepackage{slashed}
\usepackage{cancel}
\usepackage{caption}
\usepackage{subcaption}
\usepackage{booktabs}
\usepackage{physics}
\usepackage{soul}

\usepackage{pifont}

\newcommand{\uv}{\text{uv}}
\newcommand{\vuv}{v_\text{uv}}
\newcommand{\ir}{\text{ir}}
\newcommand{\rir}{r_\text{ir}}
\newcommand{\alphair}{\alpha_\text{ir}}

\usepackage{feynmp}
\usepackage{graphicx}
\usepackage{slashed}
\usepackage{subcaption}

\usepackage[bottom]{footmisc} 

\title{Consequences of a Stabilizing Field's Self-Interactions for RS Cosmology}
\author{Rashmish K.~Mishra and}
\author{Lisa Randall}

\affiliation{Harvard University, 17 Oxford Street, Cambridge, MA, 02139, USA}

\abstract{
It has been argued that the Randall-Sundrum (RS) phase transition rate is suppressed when the holographic theory corresponds to a large $N$ Yang-Mills and when the stabilizing field has a small mass. Here we argue that self-interactions can alleviate the latter suppression.
We consider a cubic term in the bulk potential for the Goldberger-Wise (GW) scalar that is responsible for stabilizing the RS geometry. Adding a cubic term suffices to separate the two roles of the GW stabilization: generating a large hierarchy and triggering confinement. We study the resulting radion potential and the dynamics of the early universe phase transition. For a negative coefficient of the cubic term, the effect of the cubic becomes important in the infra-red, and the resulting radion potential is deeper, thereby increasing the radion mass while maintaining a large hierarchy. Staying within the radion effective field theory, we calculate the rate of bubble nucleation from the hot phase to the confined RS phase, both in thin and thick wall limits. The cubic term enhances the rate and allows relaxing the condition on the maximum number of colors $N_\text{max}$ of the dual theory for which the phase transition can be completed. Importantly, this reduces the amount of supercooling that the false vacuum undergoes, increases the peak frequency of the gravitational waves (GW) produced from bubble collisions, and reduces the strength of the GW signal. The reduced GW signal is however still within the reach of proposed space-based GW detectors.
}

\begin{document} 
\maketitle
\flushbottom

\section{Introduction}
\label{sec:Intro}
The Randall-Sundrum (RS) framework based on warped extra-dimensional geometry is an elegant way to understand various hierarchies in the Standard Model (SM), and provides a rich framework for exploring several directions in both formal and phenomenological research~\cite{Agrawal:2022rqd}. The early cosmological history of these models is a confluence of many interesting phenomena. At low temperatures, the RS phase is the thermodynamically stable phase. At high temperatures, the RS phase is only metastable, while the stable phase is given by a black-brane geometry, which in the dual field theory corresponds to the deconfined phase. In minimal constructions, these two phases are separated by a barrier~\cite{Creminelli:2001th}, and a transition between the two phases proceeds by bubble nucleation. 

Starting in the deconfined phase at high temperatures, the rate of transition to the confined phase is very suppressed in the minimal models, and they generically supercool past the critical temperature. The field configuration that allows tunneling from one phase to another is a gravitational instanton, the grammar for which is an active area of investigation. Early cosmological history of RS models is therefore tied to the physics of confinement, supercooling, and gravitational instantons. All these effects have direct phenomenological relevance---the confinement/deconfinement phase transition and its order is relevant for potential gravitational wave (GW) signals.
Supercooling is important for estimating the peak frequency and abundance for the GW signal~\cite{Randall:2006py, Konstandin:2011dr}, the present abundance of relics from that era, as well as when and whether the phase transition completes.\footnote{See ref.~\cite{Levi:2022bzt} for a general discussion of supercooling at both weak and strong coupling.} The characteristic peak frequency and the frequency dependence of the GW abundance have an imprint of the energy scales involved and can be probed in present and proposed GW detectors~\cite{LISA:2017pwj, Baker:2019nia, Seto:2001qf, Kawamura:2011zz, Yagi:2011wg, Isoyama:2018rjb, Crowder:2005nr, Corbin:2005ny, Harry:2006fi}. Finally, the exploration of gravitational instantons is relevant for all of these since it affects the rate of transition. These considerations make an exploration of the theoretical and phenomenological aspects of phase transitions in RS framework a well-motivated direction to pursue.  If the phase transition is first-order, the resulting GWs can provide access to the cosmological history of the universe~\cite{Kosowsky:1991ua, Kosowsky:1992vn, Kosowsky:1992rz, Kamionkowski:1993fg} and point towards yet to be discovered beyond the Standard Model (BSM) physics. 

In a realistic UV complete warped scenario, we generically expect significant IR modifications, since in the dual picture, the theory is close to confinement, and is very far away from a conformal field theory (CFT). With this motivation, in the present work we allow for an IR modification in the RS framework stabilized by a Goldberger-Wise (GW) scalar. We focus on how such a modification affects various cosmological features.

More concretely, we consider a quadratic and a cubic term in the bulk potential for the stabilizing GW field, with both their coefficients negative, so that the GW profile grows in the IR, and the effect of the cubic becomes important in the IR as well. Generically one expects even higher-order terms in the potential, but for the present purposes, a cubic suffices to model the IR modification. A non-zero cubic allows splitting the two roles played by the GW mechanism---a logarithmic running which gives a large hierarchy, and the triggering of the IR brane. As we will see, a non-zero cubic term can change the shape of the radion potential while maintaining a large hierarchy. This translates to a modification of the free energy and the bounce action for the phase transition. In the present work, we will focus on the modifications on the RS phase only and stay in the regime where the backreaction is not important.

A suppressed rate of phase transition in the minimal models can be tracked down to two parametric reasons: large $N_c$ and small $\delta$. Here $N_c$ is the number of colors in the dual theory, and $\delta$ characterizes CFT breaking in the IR where the phase transition takes place (and is a function of the parameters of the stabilization mechanism).  For theoretical control, we need $N_c \gg 1$. The minimal RS models, stabilized with a quadratic bulk GW potential, also have $\delta \ll 1$. The present work can be understood as a way to enhance the rate by increasing $\delta$, and is similar in spirit to other such attempts in the literature. Addressing the $N_c$ suppression directly will require a better modeling of the IR dynamics, and including backreaction, which we will present in a future work.

Cosmological aspects of RS models have been studied both in the minimal scenario~\cite{Creminelli:2001th}, and with interesting variations~\cite{Nardini:2007me, Konstandin:2010cd, Konstandin:2011dr, vonHarling:2017yew, Dillon:2017ctw, Bunk:2017fic, Bruggisser:2018mrt, Baratella:2018pxi, Megias:2018sxv, Pomarol:2019aae, Agashe:2019lhy, Fujikura:2019oyi, Megias:2020vek, Bigazzi:2020phm, Agashe:2020lfz, Agrawal:2021alq, Csaki:2023pwy, Girmohanta:2023sjv, Eroncel:2023uqf} that were aimed towards addressing the issues in the minimal scenario. The authors in ref.~\cite{Creminelli:2001th} pointed out that without a stabilization mechanism, if the RS model is in the deconfined phase at high temperatures, it never completes a transition to the confined phase. With a stabilization mechanism from a GW scalar, the rate for the phase transition is extremely suppressed, and is very constraining on what constitute viable physical parameters. Ref.~\cite{Konstandin:2010cd} pointed out that by including back-reaction, the situation is less constraining than originally claimed. In subsequent work, ref.~\cite{vonHarling:2017yew} used corrections to the radion potential coming from the QCD condensate, which reduced the barrier depth, and hence enhanced the rate for the phase transition. The authors in~\cite{Baratella:2018pxi} further required the 4D CFT degrees of freedom to flow to another fixed point after QCD confinement, thereby gaining more theoretical control, and gave a geometrical picture for this situation where the mass of the GW scalar is tachyonic~\cite{Pomarol:2019aae}. The authors in~\cite{Agashe:2019lhy,Agashe:2020lfz} used similar ingredients where they modeled the IR flow to another fixed point by a special form of bulk potential for the GW field. In an orthogonal direction, the authors of ref.~\cite{Agrawal:2021alq} used finite temperature corrections to the combined potential of radion and other light fields to address the issue. Ref.~\cite{Csaki:2023pwy} used a relevant UV deformation to model the CFT breaking in the IR, which resulted in a deeper radion potential. See also~\cite{Girmohanta:2023sjv,Eroncel:2023uqf} for other interesting alternatives. 

Compared to the previous approaches, the present work is different in the following ways: the modified bulk potential we consider is generic, and is expected in a realistic UV completion. We stay within the radion EFT parameter space. We use a combination of analytical and numerical methods to handle the computations. We consider both thin and thick wall limits, for completeness. Our approach is a first step towards systematically including strongly coupled IR effects. In some respects our results mimic those of \cite{Agashe:2019lhy,Agashe:2020lfz}, who explicitly separated the two roles of the radion by having one CFT to establish the hierarchy and the second to establish the mass. Though perhaps under less control, our model is more generic and reproduces the second feature organically. Once the GW field is sufficiently big, playing a role in triggering the IR phase transition, it also contributes to the radion mass and eases the phase transition completion.

The outline of the paper is the following. In sec.~\ref{sec:ActionAndScalarProfiles} we set up the notation and obtain the GW scalar profile for both phases. In sec.~\ref{sec:RadPot} we derive the radion potential, and in sec.~\ref{sec:FE-of-phases} we derive the free energy of the two phases as a function of temperature. In sec.~\ref{sec:DynamicsOfPhaseTransition} we calculate the rate of phase transition, first in thin-wall approximation and then away from it, focusing on the role of the IR effects. Sec.~\ref{sec:Results} contains the results, and a conclusion follows in sec.~\ref{sec:GeneralComments}. Technical details of the calculations are presented in app.~\ref{app:ApproxSols}, \ref{app:RadionPotential}, \ref{app:BounceAction}.

\section{5D action and scalar profiles}
\label{sec:ActionAndScalarProfiles}
To set the stage we first fix some notation. We consider 5D spacetime with locally constant negative cosmological constant. The general solution can be parameterized as
\begin{align}
    \dd s^2 &= - e^{-2r}\left(1-e^{4(r-r_h)}\right)\,\dd t^2 + e^{-2r} \dd\vec{x}^2 + \frac{\dd r^2}{1-e^{4(r-r_h)}}\: ,
    \label{eq:RSandAdSSmetric}
\end{align}
which is the metric for Anti-de Sitter Schwarzschild (AdSS) space, with the asymptotic boundary of AdS space at $r \to -\infty$ and a horizon $r=r_h$, extended in the transverse directions (a black brane). We have set the AdS scale $\ell_\text{AdS} = 1$ here, and in what follows. In the limit of $r_h \to \infty$, this reduces to the Poincare patch of AdS space. 

A UV brane with appropriate (positive) tension, whose location can be chosen to be at $r_\text{uv} = 0$, cuts off the UV region of the spacetime and ensures a normalizable 4d graviton. For the $r_h=\infty$ case, the spacetime will be truncated at the IR brane located at $r = r_\text{ir}$, again with an appropriately chosen (negative) tension. We therefore have two spacetime metrics to consider, which we refer to as the BB (Black Brane) and RS (Randall Sundrum) spacetimes respectively: 
\begin{align}
    \text{RS} &:\:\: 
    \dd s^2 = - e^{-2r}\,\dd t^2 + e^{-2r} \dd\vec{x}^2 + \dd r^2\: , \qquad\qquad\qquad\qquad\qquad\:\: 0 \leq r \leq r_\ir\:,
    \nonumber \\
    \text{BB} &:\:\: 
    \dd s^2 = - e^{-2r}\left(1-e^{4(r-r_h)}\right)\,\dd t^2 + e^{-2r} \dd\vec{x}^2 + \frac{\dd r^2}{1-e^{4(r-r_h)}}\: , \:\:0 \leq r \leq r_h \:.
\end{align}
The RS and BB spacetimes are dual to the confined and the deconfined phases respectively in the field theory.
The Hawking radiation from the horizon in the BB geometry gives a temperature to the black hole, which is a function of the location of the horizon. The effect of a finite temperature $T$ can be studied in the Euclidean version of the spacetime, with the Euclidean time $t_E$ identified with a period $\beta = 1/T$. For $r_h \neq 0$, the Euclidean continuation of eq.~\eqref{eq:RSandAdSSmetric} with a periodic $t_E$ is smooth only when 
\begin{align}
    \beta = \pi \exp(r_h) \:,
    \label{eq:Temperature-vs-horizon}
\end{align} 
whereas any $\beta$ is fine for $r_h=\infty$ (RS background). Since we need to consider the dynamics at finite temperature, we will work in Euclidean compactified time.

The location of the IR brane is a modulus in the RS geometry, and needs a stabilization mechanism to have a fixed value. Such a stabilization can be provided by a 5D GW scalar $\chi$ with appropriately chosen parameters such that it gets a profile along the extra dimension, and generates a potential for the field $r_\text{ir}(x)$. In this work we consider more general potentials and boundary conditions than the original proposal~\cite{Goldberger:1999uk,Goldberger:1999un}, and argue that these modifications are well-motivated and geared towards modeling the IR dynamics appropriately. Specifically, we consider a 5D scalar $\chi$, with the action
\begin{align}
    S_\chi &= \int d^5x \sqrt{g}\left(-\frac12\left(\partial\,\chi\right)^2-V_B(\chi)\right) - \sum_{i}\int d^4x\,\sqrt{g_i}\,V_i(\chi) \: ,
    \label{eq:GW-action}
\end{align}
with a bulk potential $V_B(\chi)$ and boundary potential(s) $V_i(\chi)$ which set the boundary conditions. Note that $i$ takes the value $(uv, ir)$ for the RS case, and only $(uv)$ for the BB case. We would like to solve for the profile of $\chi$ in the two backgrounds, and in the limit of small back-reaction.

The choice of the bulk and brane localized potentials is governed by the dynamics one wants to model. In the dual interpretation, the bulk GW potential can be mapped to the renormalization flow of a deformation. For the effect to become more important in the IR, the deformation should be relevant, and higher-order terms in the beta function should become important as the deformation grows. A constant mass for the GW scalar corresponds to a constant beta function. Higher order terms in the beta function correspond to higher order terms in the GW potential~\cite{Rattazzi:2000hs,Chacko:2013dra}. We add a cubic term in the bulk GW potential, in addition to the mass term, to model the higher-order terms in the beta function. With this motivation, we choose the bulk potential to be
\begin{align}
    V_B(\chi) &= 
    2\epsilon_2\chi^2+\frac43\epsilon_3\chi^3\:,
    \label{eq:GW-bulkPotential}
\end{align}
with $\epsilon_2 < 0, \epsilon_3 <0$. The first sign, $\epsilon_2 < 0$, corresponds to a relevant deformation of the dual theory, and ensures a logarithmic running for small $|\epsilon_2|$. The second sign, $\epsilon_3 < 0$, ensures the deformation gets larger in the IR. We have kept to a cubic term in the potential, which is sufficient to capture the effect of strong coupling as we show later, although there can in general be quartic and higher-order terms as well.\footnote{Ref.~\cite{Agashe:2020lfz} used a GW potential with up to quartic terms to model a specific flow in the dual theory, from a UV fixed point to an IR fixed point. The dynamics we want to model in the present work is different.}

The mass term and the self-interaction term in the GW potential model different aspects of the dynamics. A small mass allows a large running, whereas the self-interaction term, which for small $\epsilon_3$ is important only in the IR, models the existence of a more complicated radion potential in the IR after confinement. It is important to separate these two effects, and study the resulting effect on radion potential and the phase transition, which is the motivation for this work.

For simplicity we choose a UV brane potential that fixes the value of $\chi$ at the UV brane. The IR brane potential is chosen to allow the scalar to adjust its value, and again for simplicity we fix its derivative. These features can be modeled by the potentials 
\begin{align}
    V_\uv(\chi) &= \beta_\uv(\chi-v_\uv)^2\:,\;\:\beta_\uv\to\infty\:,
    \nonumber \\
    V_\ir(\chi) &= 2\alphair\chi\:.
    \label{eq:GW-branePotential}
\end{align}
Note that these are simplifications, and a more complete analysis should allow for mixed boundary conditions at both the UV and the IR. Also note that the IR boundary condition is not relevant for the BB case---the GW profile is required to have a ``regular'' behavior at the horizon. For a truly holographic interpretation, we would not necessarily have a boundary condition imposed directly in the IR but we follow the original analysis and do this for simplicity.

Given the bulk and boundary potentials, the GW field develops a profile $\chi(r)$ along the radial direction, which is different for the RS and BB backgrounds because the IR boundary condition is different for the two cases. While an exact solution can be obtained numerically, it is possible to get approximate analytical solutions by dividing the bulk into different parts where different terms in the equation dominate. Leaving the details to appendix~\ref{app:ApproxSols}, the leading order solutions are given as
\begin{align}
    \chi_\text{RS}(r) &= -\frac{\alpha_\ir}{4} e^{4(r-r_\ir)} + \frac{v_\uv e^{-\epsilon_2 r}}{1 + v_\uv \epsilon_3 \left(\frac{1-e^{-\epsilon_2 r}}{\epsilon_2}\right)} \:, \:\: 0 \leq r \leq r_\text{ir}\:,
    \label{eq:GW-profile-RS}
    \\
    \chi_\text{BB}(r) &= \frac{v_\uv e^{-\epsilon_2 r}}{1 + v_\uv \epsilon_3 \left(\frac{1-e^{-\epsilon_2 r}}{\epsilon_2}\right)} \:, \qquad\qquad\qquad\:\:\:\:\:\: 0 \leq r \leq r_h\:.
    \label{eq:GW-profile-BB}
\end{align}
The solutions are obtained in the limit of $|\epsilon_2|\ll 1, v_\uv \ll 1, r_\text{ir} \gg 1, |\epsilon_2| r_\ir \lesssim 1, |\epsilon_3| r_\ir \lesssim 1$ ($r_\ir \to r_h$ in the corresponding conditions for the BB solution). In the case of $\epsilon_2<0$, $\chi$ becomes singular at $r_s = -(1/\epsilon_2)\log(1+\epsilon_2/v_\uv\epsilon_3)$ and is an artifact of the analytical solution failing to be a good approximation. By appropriately choosing parameters $r_s$ can be made large and as long as  $r_s \gg r_\ir, r_h$, we can trust the solutions. 

A few comments are in order about these solutions. First, for the RS solution, the term proportional to $\alpha_\ir$ can be important only very close to $r=r_\ir$. Second, the requirement of regularity of the solution in the BB background is the same as dropping the term proportional to $\alphair$ in the RS solution, at least to leading order in the approximation. Finally, these approximate solutions clarify the effect of $\epsilon_3$. For small $r$, the term proportional to $v_\uv$ can be written as $v_\uv\exp(-(\epsilon_2+v_\uv\epsilon_3)r)$. The self-interaction term therefore effectively acts as an additive effect to $\epsilon_2$.  As $r$ increases, this effect compounds.

A good diagnostic to quantify the effect of a non-zero $\epsilon_3$ is to look at the ratio of the mass of the radion to the IR scale.  The model requires a large hierarchy, which is the result of logarithmic running.  A non-zero $\epsilon_3$ allows higher order terms in the radion potential to balance in the IR.  We will have more to say about this in the next section.

\begin{figure}[h!]
    \centering
    \includegraphics[scale=0.7]{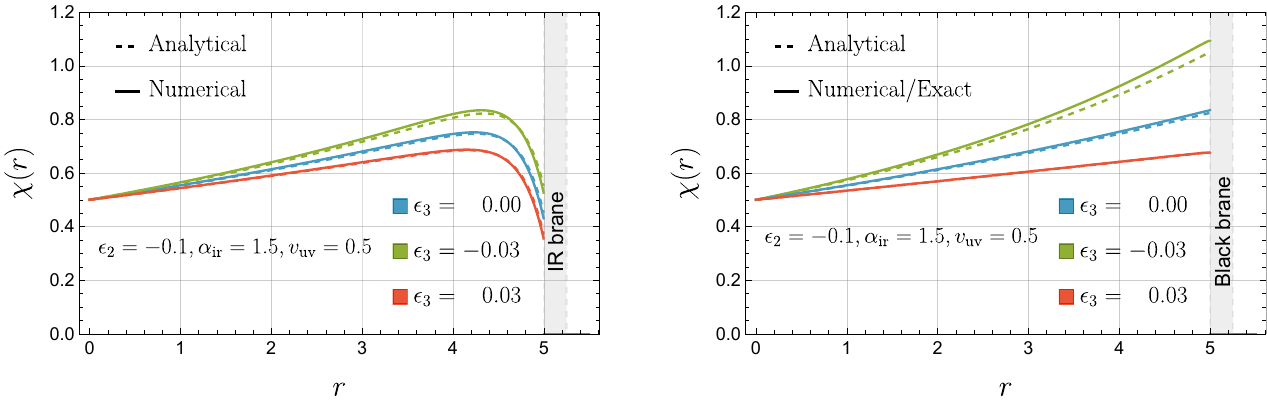}
    \caption{
    \small{GW scalar profile with and without self-interactions, for the RS background (left) and the BB background (right), for other parameters fixed. Solid lines (numerical) show good agreement with the approximate solutions (in dashed). A negative (positive) $\epsilon_3$ leads to more (less) growth of the profile in the IR.}
    }
    \label{fig:GW-profile-SI}
\end{figure}

Figure~\ref{fig:GW-profile-SI} shows a comparison between the approximate solutions in eqs.~\eqref{eq:GW-profile-RS}, \eqref{eq:GW-profile-BB} with numerical solutions, for some choices of parameters. The procedure to numerically obtain regular solutions in the BB background are discussed in appendix~\ref{app:ApproxSols}.

\section{Radion potential}
\label{sec:RadPot}
Given the profile for $\chi(r)$, we can calculate the resultant radion potential. The basic idea is that as one varies $r_\ir$, the energy contained in the potential for $\chi(r)$ changes and for a choice of parameters, a suitable minimum can be obtained. The purely gravitational part of the action gives a kinetic term for $r_\ir$. Evaluating the $\chi$ action on the approximate solution for $\chi(r)$ in eq.~\eqref{eq:GW-profile-RS}, one obtains the potential for $r_\ir$. In terms of the field $\varphi = e^{-r_\ir}$, the 4D action is given as
\begin{align}
    S&= \int d^4x \sqrt{g}\left(-12M_5^3\left(\partial\varphi\right)^2 - V(\varphi)\right)\:,
    \nonumber \\
    V(\varphi) &= 24 M_5^3\kappa^4\,\varphi^4\left(1+ \frac{a_2}{24 M_5^3\kappa^4}\frac{\lambda\varphi^{\epsilon_2}}{1-\lambda\varphi^{\epsilon_2}} - \frac{a_3}{24 M_5^3\kappa^4}\,\log (1-\lambda\varphi^{\epsilon_2})\right),\:
    \nonumber\\
    \qquad&\lambda = \frac{v_\text{uv} \epsilon_{32}}{1+ v_\text{uv}\epsilon_{32}},\:
    \epsilon_{32} = \frac{\epsilon_3}{\epsilon_2},\:a_2 = -\frac{1}{32}\epsilon_2\alphair^2-\frac{\epsilon_2}{\epsilon_3}\alphair + 2\alphair,\: a_3 = \frac12\frac{\epsilon_2}{\epsilon_3}\alphair\:.
    \label{eq:RadPotGeneral}
\end{align}
The details of the computation for $V(\varphi)$ are given in app.~\ref{app:RadionPotential}. Note that $\varphi$ is not canonically normalized in our notation. We have pulled an overall factor of $M_5^3$ outside from the potential, and the parameter $\kappa \lesssim 1$ for small back-reaction.\footnote{We are working in the glueball normalization where the quartic scales as $N_c^2 \sim M_5^3$.} In the limit of $\lambda\varphi^{\epsilon_2} \ll 1$ (note that for this we need $\lambda \sim \mathcal{O}(\vuv) \ll 1$, since $\varphi^{\epsilon_2}$ can be $\mathcal{O}(1)$, for $\epsilon_2 < 0$), the potential can be expanded in a power series:
\begin{align}
    V(\varphi) &= \varphi^4
    \left(b_0 + b_1 \lambda \varphi^{\epsilon_2}
    + b_2 \lambda^2 \varphi^{2\epsilon_2}
    + b_3 \lambda^3 \varphi^{3\epsilon_2}
    + \cdots
    \right)\:.
    \label{eq:RadPot-generic}
\end{align}
The coefficients $b_i$ are readily calculable given the explicit form of the potential, and are functions of $\alphair, \vuv, \epsilon_2$ and $\epsilon_3$. In the limit of $\epsilon_3\to0$, only $b_0$ and $b_1$ are non-zero, and the potential simplifies to the familiar racetrack form. Note that in this limit, $\lambda\to 0$ but it is balanced by the  $\epsilon_3$ in the denominator of $a_2$ and $a_3$ in eq.~\eqref{eq:RadPotGeneral}. 
\begin{align}
    V(\varphi) &\underset{\epsilon_3\to0}{=}\:\: 
    24 M_5^3 \kappa^4\,\varphi^4\left(1 - \frac{1}{48 M_5^3\kappa^4}\alphair v_\uv \varphi^{\epsilon_2}\right)
    \nonumber \\
    &= 24 M_5^3 \kappa^4\,\varphi^4\left(1 -\frac{1}{1+\epsilon/4}\left(\frac{\varphi}{\varphi_\text{min}}\right)^{\epsilon_2}\right)\:.
\end{align}
The $b_2$ and higher order terms in eq.~\eqref{eq:RadPot-generic} mimic the effects of strong dynamics---as the running coupling grows to be big, the higher order terms become important~\cite{Rattazzi:2000hs}. In fact, we will see that for a relevant choice of parameters, we have to include several terms in the expansion. The effect of higher order terms is that for a small $\varphi_\text{min}$ (large hierarchy), various combinations of terms can balance each other and give a minimum in the radion potential in which case the second derivative of the potential at the minimum can be enhanced. This is the usual expectation that for a strong breaking in the IR, the radion is not parametrically light anymore. To understand this enhancement, let's first consider the generic form of the radion potential
\begin{align}
    V(\varphi) &= b_0\,\varphi^4\,P(\varphi^{\epsilon_2})\:,
    \label{eq:RadPotGen}
\end{align}
where $P(x)$ is a polynomial in $x$ of some given order, with the first term $1$ (since we have factored out an overall constant in eq.~\eqref{eq:RadPotGen}). 
Expanding eq.~\eqref{eq:RadPotGeneral}, some of the terms in $P(x)$ are explicitly given as
\begin{align}
    b_0\,P(x) &= b_0 +\sum_{i\geq1}b_i x^i\:,
    \qquad b_0 = 24 M_5^3\kappa^4\:,
    \nonumber \\
    b_1 &= -\vuv 
    \left(\frac12 \alphair-2\alphair\frac{\epsilon_3}{\epsilon_2}+\frac{1}{32}\alphair^2\epsilon_3\right)
    \left(1+\frac{\vuv\epsilon_3}{\epsilon_2}\right)^{-1}\:,
    \nonumber \\
    b_2 &= -\vuv \left(\frac{\vuv\epsilon_3}{\epsilon_2}\right) 
    \left(\frac34 \alphair-2\alphair\frac{\epsilon_3}{\epsilon_2}+\frac{1}{32}\alphair^2\epsilon_3\right)
    \left(1+\frac{\vuv\epsilon_3}{\epsilon_2}\right)^{-2}\:,
    \nonumber \\
    b_3 &= -\vuv \left(\frac{\vuv\epsilon_3}{\epsilon_2}\right)^2 
    \left(\frac34 \alphair-2\alphair\frac{\epsilon_3}{\epsilon_2}+\frac{1}{32}\alphair^2\epsilon_3\right)
    \left(1+\frac{\vuv\epsilon_3}{\epsilon_2}\right)^{-2}\:,
    \nonumber \\
    \vdots\:\:\:
\end{align}
For $\vuv \lesssim 1$ and $\epsilon_3/\epsilon_2 \lesssim 1$, higher order terms are successively smaller. However, near the minimum, $x = \varphi^{\epsilon_2} \gtrsim 1$ so that higher powers of $x$ are bigger. Therefore, near the minimum, terms in $P(x)$ are products of successively increasing and successively decreasing factors. For certain choice of parameters, a combination of terms balance each other. Since $x\gtrsim 1$, we still get a large hierarchy. This discussion also makes it clear that for a small $\epsilon_3$,  the higher order terms can contribute without changing the hierarchy too much when $\alphair$ and $\vuv$ take larger values. At such values, $P$ is enhanced and so is the second derivative of the potential.

In summary, a non-zero $\epsilon_3$, in conjunction with other parameters of the GW sector, can increase the radion mass while generating a  similar hierarchy. To illustrate this, we will work with four benchmark parameters $\textbf{A,B,C,D}$ in table~\ref{tab:params-ABCD}, for which $\varphi_\text{min} \sim 10^{-16}$.\footnote{In table~\ref{tab:params-ABCD}, $\varphi_\text{min}$ is calculated for $M_5^3=N_c^2/16\pi^2, N_c=1$. For a different $N_c$, $\alpha_\ir$ and $v_\uv$ have to be adjusted to keep $\varphi_\text{min}$ fixed. The mass of the physical radion and $T_c$ (defined in eq.~\eqref{eq:Tc-def}) do not change with $N_c$.} These parameters are chosen with certain self-consistency conditions in mind. Requiring to stay in the radion EFT, we need to ensure that the radion mass is at most or slightly smaller than the Kaluza-Klein (KK) scale which sets the mass of other KK modes. This also ensures that the back-reaction on the geometry from the GW scalar can be ignored. Another requirement is to have $T_c/\varphi_\text{min} \lesssim 1$ so that temperature corrections to the potential can be ignored, at least in the vicinity of the minimum. $T_c$ is set by the value of the potential at the minimum $V(\varphi_\text{min})$  (as discussed in the next section).
\begin{table}
\centering
    \begin{tabular}{||c|c|c|c|c|c||c|c|c||}
        \hline 
        & $\kappa$ & $\epsilon_2$ & $\epsilon_3$ & $\alphair$ & $\vuv$ & $\varphi_\text{min} \times 10^{16}$ & $V''(\varphi_\text{min})/\varphi_\text{min}^2$ & $-V(\varphi_\text{min})/\varphi_\text{min}^4$ \\
        \hline \hline
        $\textbf{A}$ & $10^{-1/4}$ & -1/25 & 0 & 1/10 & 1/14 & 1.47 &0.002 & $10^{-4}$ \\
        $\textbf{B}$ & $10^{-1/4}$ & -1/25 & -1/100 & 5/2 & 1/14 & 1.09 &0.005 & $3 \times 10^{-4}$ \\
        $\textbf{C}$ & $10^{-1/4}$ & -1/25 & -1/90 & 5/2 & 1/5 & 0.86 &0.032 & $2 \times 10^{-3}$\\
        $\textbf{D}$ & $10^{-1/4}$ & -1/25 & -1/81 & 5/2 & 1/3 & 0.59 &0.135 & $8 \times 10^{-3}$\\
        \hline
    \end{tabular}
    \caption{\small{Benchmark choice of parameters to show the effect of self-interaction.}}
    \label{tab:params-ABCD}
\end{table} 

\begin{figure}[ht!]
    \centering
    \begin{subfigure}[b]{0.99 \textwidth}
    \centering
    \includegraphics[scale=0.85]{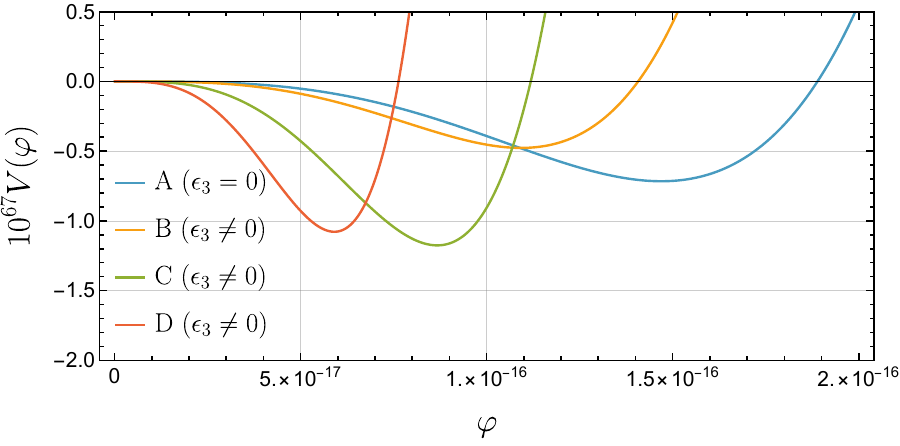}
    \caption{
        \small{Radion potentials near the minimum, for the parameter choices in table~\ref{tab:params-ABCD}. A deeper potential results from a non-zero $\epsilon_3$.}
        }
    \label{fig:RadPot-withSI-paramsABCD}
    \end{subfigure}
    \begin{subfigure}[b]{0.99 \textwidth}
    \centering
    \includegraphics[scale=0.7]{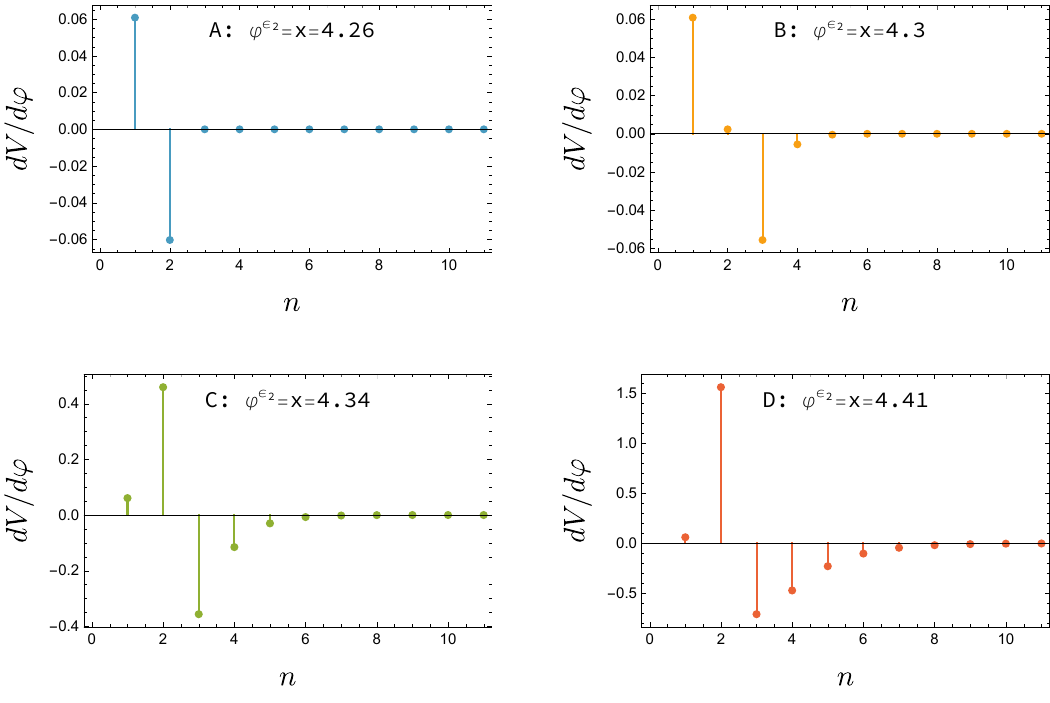}
    \caption{
        \small{Various terms in the series expansion of the derivative of the radion potential, for the parameter choices in table~\ref{tab:params-ABCD}. The terms balancing near the minimum are, top left: first and second, top right: first and third, bottom left: second and third, and bottom right: second, third and fourth.}
        }
    \label{fig:RadPot-withSI-paramsABCD-Series}
    \end{subfigure}
\caption{\small{Radion potentials, for the parameter choices in table~\ref{tab:params-ABCD}.}}
\label{fig:RadPot-withSI}
\end{figure}

Fig.~\ref{fig:RadPot-withSI-paramsABCD} shows the radion potential for these parameters. For $\textbf{B,C,D}$, with $\epsilon_3\neq0$, a deeper potential at the minimum can be clearly seen. Fig.~\ref{fig:RadPot-withSI-paramsABCD-Series} shows the value of various terms in the series expansion of the derivative of the potential near the minimum. It is clear that for $\textbf{A}$, the first and the second terms balance, for $\textbf{B}$, the first and the third terms balance, for $\textbf{C}$, the second and the third terms balance, and for $\textbf{D}$, the second term balances the third and fourth terms together. For $\textbf{C,D}$, $\vuv$ was increased to make sure the higher order terms are enhanced. Further, the magnitude of the dominant term is largest in $\textbf{D}$, which is correlated with the largest value of the second derivative at the minimum. One can estimate the second derivative at the minimum once we know which terms balance which. Starting with the $m^\text{th}$ term in the potential, $V_m = b_m\,\varphi^{4+m\epsilon_2}$, the derivative and the second derivative are
\begin{align}
    V'_m =  b_m\,(4+m\epsilon_2)\,\varphi^{3+m\epsilon_2}\:,\qquad V''_m = b_m\,(4+m\epsilon_2)(3+m\epsilon_2)\,\varphi^{2+m\epsilon_2}\:.
\end{align}
The $m^\text{th}$ and $n^\text{th}$ term in $V'$ can balance near the minimum if
\begin{align}
    b_m\,(4+m\epsilon_2)\,\varphi_\text{min}^{3+m\epsilon_2} \sim -b_n\,(4+n\epsilon_2)\,\varphi_\text{min}^{3+n\epsilon_2}\:.
\end{align} 
Considering these two terms, and using the above, the second derivative at the minimum is
\begin{align}
    V''/\varphi_\text{min}^2 \sim b_m\,(4+m\epsilon_2)(m-n)\epsilon_2\,\varphi^{m\epsilon_2} \sim b_n\,(4+n\epsilon_2)(n-m)\epsilon_2\,\varphi^{n\epsilon_2}\:.
    \label{eq:Vrad-der2-Estimate}
\end{align}
Using $b_0 = 24 M_5^3\kappa^4$, $b_{m>0} \sim \vuv \left(\vuv\epsilon_3/\epsilon_2\right)^{m-1}\alphair$ and $\varphi^{\epsilon_2} \sim (10^{-16})^{-1/25}\sim 4.3$ near the minimum, the above estimate for $V''/\varphi_\text{min}^2$ matches the numbers in table~\ref{tab:params-ABCD} obtained numerically. For parameters $\textbf{C,D}$, the dominant term at the minimum scales as $1/\epsilon_2$ (i.e. $m=2$), and this parametrically cancels the $\epsilon_2$ factor in the numerator in eq.~\eqref{eq:Vrad-der2-Estimate}.

A word of caution for the obtained radion potential: for $\epsilon_2 < 0 , \epsilon_3 < 0$, which is the case at hand, the potential has a singularity at $\varphi_s = (1/\lambda)^{1/\epsilon_2}$. The singularity is coming from the analytical solution for $\chi(r)$ breaking down, and since we used this analytical solution to obtain the radion potential we see it in the potential too. In a more complete but numerically tedious calculation, this singularity would not appear. 
The potential in~\eqref{eq:RadPotGeneral} therefore cannot be trusted for large $\rir$, or equivalently for $\varphi\to0$. 
This is to be kept in mind when we discuss the thick wall results in sec.~\ref{sec:DynamicsOfPhaseTransition} (and the computational details in app.~\ref{app:BounceAction}), which probe small values of $\varphi$ as the temperature becomes small. 
At a finite temperature $T$, the potential is expected to be modified for $\varphi \lesssim T$. As we will explain later, we use the potential only in the $\varphi \gtrsim T$ region. The expression for the radion potential is therefore useful as long as $\varphi_s \ll T$.  
This also means we cannot extend the analysis to arbitrarily low temperatures. 
As we will show, before running into this issue a significant reduction in bounce action can be achieved.

\section{Free energy of the phases}
\label{sec:FE-of-phases}
For both RS and BB phases, the free energy gets contributions from the gravitational and the GW sectors. Since we are not including back-reaction, the gravitational calculation is the same as reported in the literature~\cite{Creminelli:2001th}. The free energy is UV sensitive, so it is more sensible to talk about the difference in the free energies between the phases. At the minimum, the difference in free energies coming purely from the gravitational part of the action is given by:
\begin{align}
    \left(F_\text{GR, BB}-F_\text{GR, RS}\right)_\text{min} &= 
    -2\pi^4 M_5^3 T^4 + C = -\frac{\pi^2}{8}\,N_c^2\,T^4 + C
    \:,
    \label{eq:FE-diff-GR}
\end{align}
where $C$ is a finite constant to be determined later, and in the last equality we have used the relation between $M_5$ and the number of colors $N_c$ of the dual theory: $16 \pi^2 M_5^3 = N_c^2$. As discussed in ref.~\cite{Creminelli:2001th}, in this computation the location of the horizon $r_h$, parameterized by a temperature $T_h = \pi \exp(r_h)$, is made a dynamical variable. For generic values of $T_h$ (in the Euclidean picture), there is a conical singularity at the horizon. Regulating the singularity one gets a contribution to the free energy for $T_h \neq T$. At the minimum, $T_h = T$, the conical singularity disappears, and $T$ is related to $r_h$ by eq.~\eqref{eq:Temperature-vs-horizon}. In deriving eq.~\eqref{eq:FE-diff-GR}, $\beta$ in the RS phase is adjusted to match the geometry at the UV cutoff (see refs.~\cite{Witten:1998zw, Creminelli:2001th} for details). 

We can effectively define eq.~\eqref{eq:FE-diff-GR} to be the free energy of the BB phase at the minimum. Taking $C \equiv 2\pi^4 M_5^3 T_c^4$, the free energy of the BB phase at the minimum can be written as
\begin{align}
    F_\text{BB, min} = 2\pi^4 M_5^3 \left(T_c^4 - T^4\right)\:.
\end{align}
At the moment, $T_c$ is a parameter and the free energy of the BB phase is positive or negative depending on whether $T<T_c$ or $T>T_c$.

In the presence of a GW scalar, the free energy of the BB phase gets an additional contribution. This can be computed by first solving for the scalar profile in the BB background, which is then used to calculate the free energy. The scalar contribution is subleading compared to the purely gravitational contribution, and we will not include it here. This is self-consistent with not including back-reaction from the scalar on the BB geometry. 

Similarly, the free energy of the RS phase gets an additional contribution in the presence of a GW scalar. Staying within the radion EFT, and at low enough temperatures, the free energy is given by the radion potential itself. The radion potential is normalized so that it vanishes at the minimum. This amounts to tuning the cosmological constant in the 4D EFT to zero. 
We therefore have
\begin{align}
    F_\text{GW, RS} = V(\varphi) - V(\varphi_\text{min})\:,\:\: F_\text{GW, RS, min} = 0\:.
\end{align}
We further require that the free energies of the two phases are equal in the $\varphi\to0$ and $T\to 0$ limits, which gives
\begin{align}
    -V(\varphi_\text{min}) = 2 \pi^4 M_5^3 T_c^4\:.
    \label{eq:Tc-def}
\end{align}
This fixes $T_c$ in terms of the radion potential. Putting everything together, we have
\begin{align}
    F_\text{BB, min}  &= 2\pi^4 M_5^3 \left(T_c^4-T^4\right)\:,
    \label{eq:FE-BB}
    \\
    F_\text{RS} &=   V(\varphi) - V(\varphi_\text{min})\:,
    \label{eq:FE-RS}
    \\
    F_\text{RS, min} &= 0\:,
    \label{eq:FE-RS-min}
    \\
    F_\text{BB, min} - F_\text{RS, min} &= 2\pi^4 M_5^3 \left(T_c^4 - T^4 \right) \:.
    \label{eq:FE-BB-minus-RS}
\end{align}
The meaning of $T_c$ is made clear by eq.~\eqref{eq:FE-BB-minus-RS}: At the critical temperature $T= T_c$, the difference in the free energy of the two phases is zero. For $T> T_c$, the BB phase has a lower free energy than the RS phase and is the preferred phase thermodynamically. As the temperature drops below $T_c$, the RS phase has a lower free energy and becomes the thermodynamically preferred phase.  

\section{Dynamics of the phase transition}
\label{sec:DynamicsOfPhaseTransition}
We assume that after the end of inflation and reheating, the RS model is in the BB phase, which is the thermodynamically stable phase at very high temperatures. As the temperature drops, the RS phase becomes thermodynamically favorable. Since both these phases are local minima of free energy, they are separated by some kind of barrier in the field space, and the phase transition is first order, proceeding by bubbles of true vacuum nucleating inside the false vacuum. The standard prescription to calculate the rate of such a phase transition~\cite{Coleman:1977py, Callan:1977pt} can be applied, but with important modifications pointed out in ref.~\cite{Creminelli:2001th}, which we mention next. 

To calculate the tunneling rate across a barrier, one has to identify the relevant fields that interpolate between the true and the false vacuum. In the present case, staying within the radion EFT, the relevant field for the tunneling rate calculation on the RS side is the radion itself. The radion is a composite, and a different field has to be identified in the BB phase. Ref~\cite{Creminelli:2001th} assumed the relevant degree of freedom in the BB phase to be $T_h$ (temperature of the horizon), for which a potential can be written, but the kinetic term is not known (see however ref.~\cite{Bigazzi:2020phm} for a discussion of this aspect). The way out was provided by noticing that the path in the field space that interpolates between the false and the true vacua has some features in the phenomenological cases of interest. One can split it into three regions: $i)$ the BB region, $ii)$ the $\varphi \lesssim T$ region, and $iii)$ the $T \lesssim \varphi \lesssim \varphi_\text{min}$ region. For $T \lesssim T_c$ (i.e. in the thin wall limit), the calculable contribution from region $iii)$ is parametrically larger than the combined contribution from $i)$ and $ii)$, so that it provides a useful estimate for the full action. As $T/T_c$ reduces, the contribution from region $iii)$ also reduces.
This means that if at a given $T/T_c$, the contribution from region $iii)$ is too 
small, one cannot ignore the other two contributions. Generically, we can take the other regions to contribute $\mathcal{O}(1)$ amount to the action, so that the actually computable action, from region $iii)$, can only be trusted for values of $T/T_c$ such that it is at least $\mathcal{O}(1)$. All in all, this means the above approach cannot be extended to very small values of $T/T_c$. Also, note that the radion potential has a singularity at $\varphi_s = (1/\lambda)^{1/\epsilon_2} \ll \varphi_\text{min}$ coming from a breakdown of the approximations that were used to derive it. Since $\varphi \gtrsim T$ in region $iii)$, we do not need to worry about this as long as $T$ is not too small. With these considerations in mind, we now calculate the bounce action in both the thin and thick wall limits

We leave the technical details of the bounce action calculation for thin and thick wall cases to app.~\ref{app:BounceAction}, and give the final expressions here. For the thin wall case, the bounce action is 
\begin{align}
    S_b &= \frac{4 }{3 \pi^7M_5^6} 
    \left(\frac{S_1}{T_c^3}\right)^3
    \frac{T_c/T}{\left(1-T^4/T_c^4\right)^2}\:,
    \nonumber \\
    S_1 &= \sqrt{48 M_5^3}\int_T^{\varphi_\text{min}} \dd \varphi \sqrt{V(\varphi)-V(\varphi_\text{min})}\:.
\end{align}
For the thick wall case, the approach is numerical. We have to minimize the action
\begin{align}
    S_b & = \int \dd^4 x \left(12 M_5^2(\partial\varphi)^2 + V(\varphi)+2\pi^4M_5^3T^4\right)\:,
\end{align}
subject to the boundary conditions $\varphi'(0)= 0$ and $\varphi'(\varphi \approx 0) = - (\pi^2/\sqrt{6}) T^2$. The second condition comes from equating the energy across the bubble boundary~\cite{Agashe:2020lfz}. Working with rescaled quantities
\begin{align}
    &\frac{1}{24M_5^3}\widetilde{V}(\varphi) \equiv \kappa^4\, \varphi^4\, v(\varphi)\:,
    \:y = \kappa\, \varphi\, T^{-1}\:,\:\: x = \kappa\, r \, T\:,
\end{align}
the thick wall bounce action looks like
\begin{align}
    S_b & = \frac{96 \pi M_5^3}{\kappa^3} \int \dd x \, x^2 \left(\frac12\left(\frac{\dd y}{\dd x}\right)^2 
    +
    y^4 \, v(T y/\kappa)
    +
    \frac{\pi^4}{12}
    \right)\:.
\end{align}
The overall factor of $M_5^3\sim N_c^2$ makes the $N_c^2$ dependence of the bounce action manifest. 

For the $O(3)$ symmetric bubbles with bounce action $S_b$, the tunnelling probability per unit time per unit 3-volume, at temperature $T$
is given by
\begin{align}
    \Gamma (T) = T^4 e^{-S_b(T)}\:,
    \label{eq:Gamma-inTermsOfSb}
\end{align}
where we have ignored constant $\mathcal{O}(1)$ multiplicative factors in the above. For the phase transition to complete, we need the probability in a Hubble volume to be at least of order 1, which translates to
\begin{align}
    \Gamma(T) \gtrsim H^4\:,
    \label{eq:Gamma-vs-Hubble}
\end{align}
where $H$ is the Hubble constant in the BB phase, and is fixed by the Friedmann equations 
\begin{align}
    H^2 = \frac{\rho_\text{total}}{3 M_\text{pl}^2}\:.
\end{align}
The energy density $\rho_\text{total}$ gets contributions from the vacuum energy and from the radiation (since the false vacuum is at a temperature $T$). Recall that the vacuum energy is tuned to zero at the RS minimum, so that when BB phase is meta-stable, it has higher free energy than the RS phase, and therefore has a positive vacuum energy. This is given as
\begin{align}
    \rho_\text{vac, BB} &= 2\pi^4M_5^3 T_c^4 
    \:.
    \label{eq:rho-vac-BB}
\end{align}
The energy density from radiation scales as $T^4$ and quickly becomes subdominant as $T\lesssim T_c$, which is necessary for the RS phase to become stable. The condition for the phase transition to complete becomes
\begin{align}
    T^4 e^{-S_b(T)} > \frac{4\pi^8}{9}\frac{M_5^6T_c^8}{M_\text{Pl}^4}
    \:,
\end{align}
or equivalently
\begin{align}
    S_b (T) < -\log \left(\frac{4\pi^8}{9}\frac{M_5^6T_c^8}{T^4 M_\text{Pl}^4} 
    \right) \equiv S_b^\text{max} (T/T_c) \:.
\end{align}
Defining the nucleation temperature $T_n$ as the temperature at which $\Gamma (T_n) = H^4$,  we get
\begin{align}
    S_b (T_n) =  S_b^\text{max} (T_n/T_c)\:.
\end{align}
Ignoring order 1 factors, for $T_c$ around the TeV scale, $S_b^\text{max} \sim 140$ and changes slowly as a function of $T/T_c$. Using $M_\text{pl}^2 = M_5^3$ ($\ell_\text{AdS} = 1$), $S_b^\text{max}$ is independent of the number of colors $N_c$. The bounce action $S_b(T)$ on the other hand, scales as $N_c^2$. Using this
we can effectively obtain a bound on the maximum $N_c$ that allows a phase transition to complete, at a given $T_n/T_c$ (i.e. the amount of supercooling):
\begin{align}
    N_\text{max} (T_n/T_c) = \sqrt{\frac{S_b^\text{max}(T_n/T_c)}{S_{b, N_c=1}(T_n)}}\:.
    \label{eq:Nmax}
\end{align}

In the next section, for the parameter choices in table~\ref{tab:params-ABCD}, we present the results for $S_b$ and $S_b^\text{max}$, both in the thin wall limit and away from it. We find that the bounce action is reduced for some of the parameter choices, and is correlated with increasing the radion mass.

\section{Results}
\label{sec:Results}
\subsection{Bounce action and maximum $N$}
\begin{figure}[ht!]
    \centering
    \begin{subfigure}[b]{0.99 \textwidth}
        \centering
        \includegraphics[width=\textwidth]{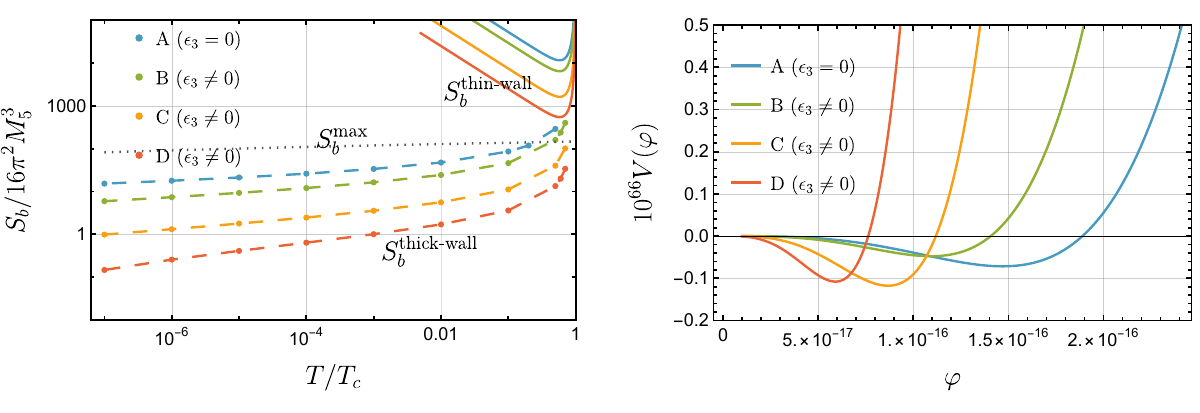}
        \caption{\small{Left: The bounce action normalized by $16\pi^2 M_5^3 = N_c^2$, for the parameter choices in table~\ref{tab:params-ABCD}. Both thin wall (solid lines) and thick wall (dashed lines) estimates for the bounce action are shown. Also shown in dotted is the maximum value of bounce action $S_b^\text{max}$, beyond which the phase transition rate to too small to compete with the Hubble expansion. Right: The corresponding radion potential for the parameter choices in table~\ref{tab:params-ABCD}.}}
        \label{fig:Thick-Thin-And-Potential-e3-Dialed-phi-min-fixed}
    \end{subfigure}
    \par\bigskip\bigskip
    \begin{subfigure}[b]{0.7 \textwidth}
        \centering
        \includegraphics[width=\textwidth]{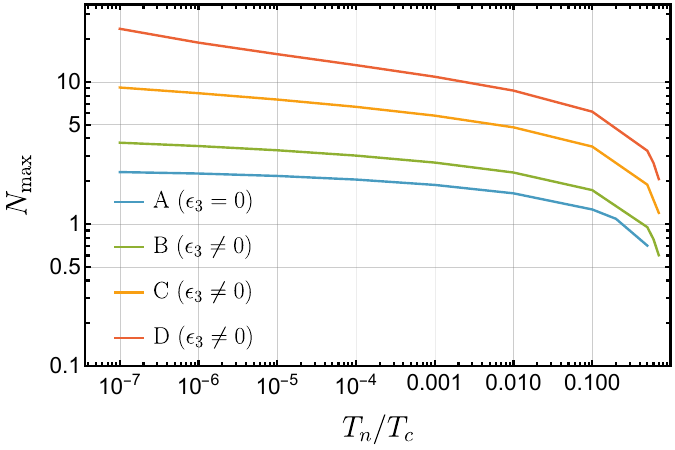}
        \caption{\small{Maximum number of colors for the phase transition to complete, as a function of $T_n/T_c$, for the parameter choice in table~\ref{tab:params-ABCD}.}}
        \label{fig:Nmax-fixedHierarchy}
    \end{subfigure}
\caption{\small{Bounce action, potential, and the maximum number of colors for the phase transition to complete, for the parameters in table~\ref{tab:params-ABCD}.}}
\label{fig:Sb-RadPot-Nmax-phiFixed}
\end{figure}

The left panel of fig.~\ref{fig:Thick-Thin-And-Potential-e3-Dialed-phi-min-fixed} shows the bounce action (normalized by $16\pi^2 M_5^3= N_c^2$) in the thin wall limit (solid lines) and away from it (dashed lines), for the parameter choices in table~\ref{tab:params-ABCD}. Also shown is the maximum bounce action (dotted) beyond which the rate is too small to compete with the Hubble expansion. We note a couple of things in fig.~\ref{fig:Thick-Thin-And-Potential-e3-Dialed-phi-min-fixed}. The thin wall curves are only applicable for $T/T_c \lesssim 1$, and for all the parameter choices, they are above the $S_b^\text{max}$ line, so that the phase transition rate is too small even for $N_c =1$, in the thin-wall limit. The phase transition completes in the thick-wall case, at a temperature $T_n$ where the $S_b$ and $S_b^\text{max}$ curves intersect (which in turn depends on $N_c$ because $S_b$ scales as $N_c^2$). For a non-zero $\epsilon_3$, the thick wall bounce action is smaller than the $\epsilon_3 = 0$ case (e.g. red vs blue curves in fig.~\ref{fig:Thick-Thin-And-Potential-e3-Dialed-phi-min-fixed}), and is a slowly varying function of $T/T_c$. Further, the variation in the thick wall bounce action as a function of $T/T_c$ is larger in the presence of a non-zero $\epsilon_3$, which has consequences for GW signal, as we explain later. 
For convenience, we also show the corresponding radion potential in the right panel of fig.~\ref{fig:Thick-Thin-And-Potential-e3-Dialed-phi-min-fixed}. We can notice a correlation between the second derivative of the radion potential at the minimum and the bounce action. When the second derivative is larger, i.e. the physical radion is heavier, the potential is deeper and the bounce action is lower. For a given amount of supercooling, say $T/T_c = 10^{-3}$, the bounce action can be lowered approximately by a factor of 30 due to a non-zero $\epsilon_3$ (e.g. comparing the red and the blue curves in fig.~\ref{fig:Thick-Thin-And-Potential-e3-Dialed-phi-min-fixed}). 

Using eq.~\eqref{eq:Nmax}, we can calculate the maximum color $N_\text{max}$ in the dual theory for which the phase transition can complete for a given amount of supercooling. Figure~\ref{fig:Nmax-fixedHierarchy} shows $N_\text{max}$ as a function of $T_n/T_c$. For some amount of supercooling, one can have $N_\text{max} \sim 10$ (e.g. the orange and red curves in fig.~\ref{fig:Nmax-fixedHierarchy}). Such values of $N_\text{max}$ are obtained at $T_n/T_c \sim 10^{-6}$, where the bounce action is becoming smaller than $\mathcal{O}(1)$ (e.g. the orange and red curves in Fig.~\ref{fig:Thick-Thin-And-Potential-e3-Dialed-phi-min-fixed}). $N_\text{max} \sim 10$ is therefore the most one can hope in the present analysis, because we have only computed a part of the bounce action that was supposed to be dominant. Once this calculable part is reduced beyond $\mathcal{O}(1)$ values, one cannot use this as a complete answer. Taking the lowest bounce action to be $\mathcal{O}(1)$, eq.~\eqref{eq:Nmax} gives $N_\text{max} \sim \sqrt{S_b^\text{max}} \sim 12$ for $S_b^\text{max} \sim 140$.

The amount of supercooling that the metastable phase experiences has important phenomenological implications. During this period of supercooling, the universe inflates, and this dilutes any matter abundances generated before the phase transition. If the framework is to address dark matter abundance and baryon asymmetry with sufficient supercooling, they must be generated after the phase transition completes. Supercooling also has implications for primordial black hole generation mechanisms, cosmological aspects of axion model building, topological defects and so on (e.g. see ~\cite{Pomarol:2019aae, Gouttenoire:2023naa} and references therein). A key more generic observation is that bubble collision at the end of the phase transition can generate gravitational waves, whose frequency and abundance are dependent on the amount of supercooling. 
For all these applications but especially with an eye to the latter,  we estimate the peak frequency and the GW abundance from bubble collisions in the next subsection. 

\subsection{Gravitational wave signal}
First-order phase transitions can give rise to stochastic gravitational wave signals, which can potentially be detected in ground and space-based GW detectors, according to their characteristic frequency and abundance (see refs.~\cite{Caprini:2015zlo, Caprini:2019egz} for a review). There are three sources of GW production in a first-order phase transition---bubble collision, sound waves, and turbulence in the plasma. The last two sources require detailed numerical analysis; here we focus on the bubble collision as a source of stochastic GW production. 

\begin{figure}[ht!]
    \centering
    \includegraphics[width=0.7\textwidth]{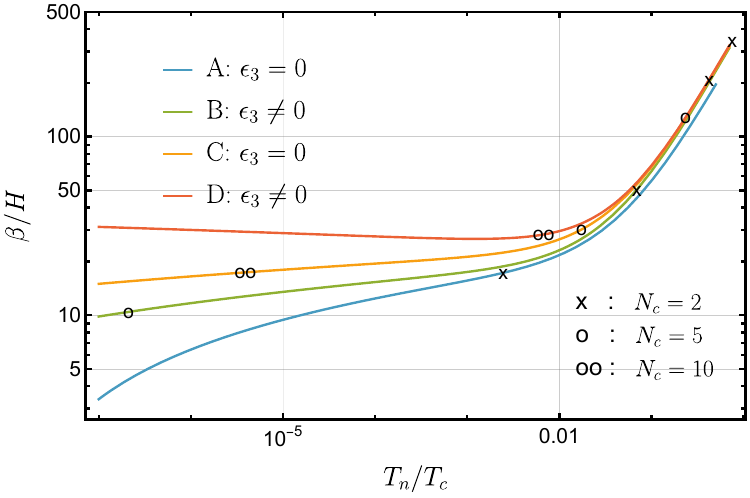}
    \caption{\small{The parameter $\beta/H$ as a function of $T_n/T_c$ for the parameter choices in table~\ref{tab:params-ABCD}. On each curve, $N$ varies as a parameter. Points with $N_c = 2, 5, 10$ are shown by markers.}}
    \label{fig:beta-vs-TnTc}
\end{figure}

The signal in GW can be characterized by the peak frequency $f_p$, and the frequency dependence of the fractional abundance $\Omega_\text{GW}\,h^2$. For the GWs generated by bubble collision, these two quantities are given by~\cite{Caprini:2015zlo}
\begin{align}
    f_p &= 0.037 \text{ mHz }
    \:\Bigg(\frac{\beta}{H}\Bigg)
    \:\Bigg(\frac{T_*}{\text{TeV}}\Bigg)
    \:\Bigg(\frac{g_*}{100}\Bigg)^{1/6}\:,
    \nonumber \\
    \Omega_\text{GW} \, h^2 (f) &= 1.3 \times 10^{-6}\:
    \:\Bigg(\frac{H}{\beta}\Bigg)^2
    \:\Bigg(\frac{100}{g_*}\Bigg)^{1/3}
    \frac{3.8(f/f_p)^{2.8}}{1+2.8(f/f_p)^{3.8}}
    \:.
    \label{eq:GW-frequencyAndSignalEstimates}
\end{align}
Here we have assumed that when the transferred latent heat is large compared to the energy of the surrounding plasma, all the latent heat is transferred to the bubble wall, and the bubble wall velocity is ultra-relativistic. $H$ here is the Hubble scale during the phase transition, $T_*$ is the temperature of the radiation bath right after the phase transition, and $g_*$ is the number of relativistic degrees of freedom in the plasma during the phase transition. The parameter $\beta/H$ is related to the duration of the phase transition~\cite{maggiore2018gravitational}, and can be calculated from the bounce action as:
\begin{align}
    \frac{\beta}{H} &= \left.-\frac{\dd \log \Gamma}{\dd \log T}\right|_{T = T_n} \approx -4 + \left.\frac{\dd S_b}{\dd \log T}\right|_{T= T_n}\:,
\end{align}
where we have used eq.~\eqref{eq:Gamma-inTermsOfSb} in the last equality above. Note that $\beta/H$ is a function of $T_n$, the temperature at which the phase transition can proceed, which in turn depends on $N_c$. A small $\beta/H$ decreases the peak frequency, but crucially increases the abundance. Figure~\ref{fig:beta-vs-TnTc} shows $\beta/H$ as a function of $T_n/T_c$ for the parameter choices in table~\ref{tab:params-ABCD}. On each curve, $N_c$ varies, which changes $T_n$. Values of $N_c = 2, 5, 10$ are shown on the individual curves. The behavior of $\beta/H$ with and without $\epsilon_3$ is very different, as seen for example by the blue and the red curves in fig.~\ref{fig:beta-vs-TnTc}. When $\epsilon_3 = 0$, $\beta/H$ is small and decreases as $T_n/T_c$ decreases, unlike the $\epsilon_3 \neq 0$ case.  
When the deformation in the IR is very small so that one is close to a CFT, the bounce action is weakly dependent on temperature~\cite{Agashe:2020lfz}, and $\beta/H$ is close to zero. This is different than the behavior seen for $\epsilon_3 \neq 0$. 

\begin{figure}[ht!]
    \centering
    \includegraphics[width=0.47\textwidth]{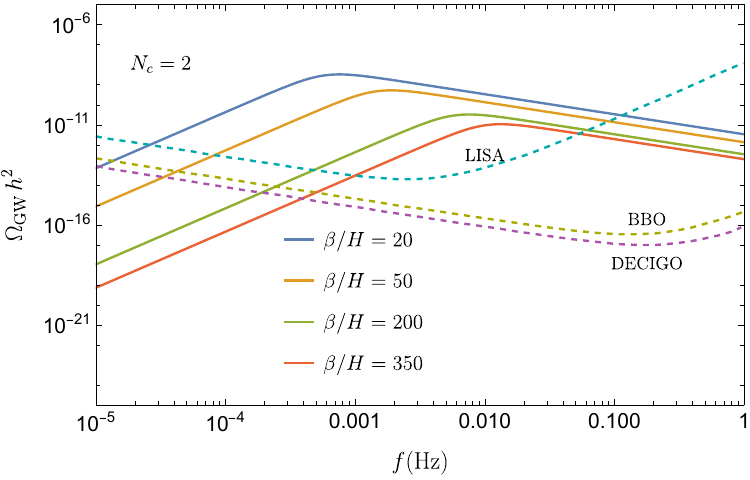}
    \includegraphics[width=0.47\textwidth]{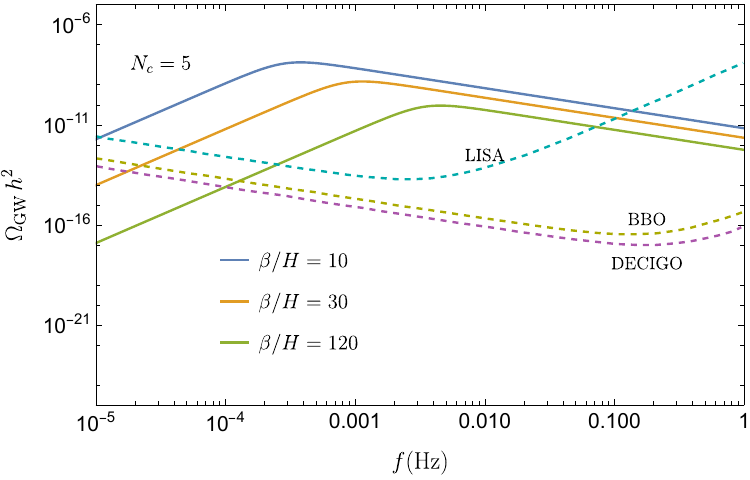}
    \caption{\small{GW abundance as a function of frequency for some values of $\beta/H$: values in the left panel correspond to $N_c = 2$ (indicated in fig.~\ref{fig:beta-vs-TnTc} by a ``x''), values in the right panel correspond to $N_c = 5$ (indicated in fig.~\ref{fig:beta-vs-TnTc} by a ``o''). Also shown is the projected reach from LISA~\cite{LISA:2017pwj, Baker:2019nia}, DECIGO~\cite{Seto:2001qf, Kawamura:2011zz, Yagi:2011wg, Isoyama:2018rjb}, and BBO~\cite{Crowder:2005nr, Corbin:2005ny, Harry:2006fi}. We have taken $g_* = 100$ and $T_* = \text{TeV}$.}}
    \label{fig:GW-signal}
\end{figure}

The increased peak frequency and reduced GW signal from a non-zero $\epsilon_3$ is still within the reach of proposed space-based GW interferometers such as LISA~\cite{LISA:2017pwj, Baker:2019nia}, DECIGO~\cite{Seto:2001qf, Kawamura:2011zz, Yagi:2011wg, Isoyama:2018rjb}, and BBO~\cite{Crowder:2005nr, Corbin:2005ny, Harry:2006fi}. In fig.~\ref{fig:GW-signal} we show the expected GW abundance as a function of frequency, for some values of $\beta/H$, and compare to the projected reach of the experiments (taking $g_* = 100$ and $T_* = \text{TeV}$ in eq.~\eqref{eq:GW-frequencyAndSignalEstimates}). For a given value of $N_c$, the values of $\beta/H$ are different for the four parameter choices in table~\ref{tab:params-ABCD}, and are at different values of $T_n/T_c$. The left panel of fig.~\ref{fig:GW-signal} shows the GW signal for $\beta/H = 20, 50, 200, 350$ (corresponding to $N_c = 2$, shown by ``x'' in fig.~\ref{fig:beta-vs-TnTc}). The value of $\beta/H$ changes significantly for a non-zero $\epsilon_3$: from $20$ ($\epsilon_3 = 0$, $T_n/T_c = 0.002$, parameter \textbf{A}) to $350$ ($\epsilon_3 = -1/81$, $T_n/T_c = 0.7$, parameter \textbf{D}). Such different values of $\beta/H$ correspond to very different peak frequencies and fractional abundances (e.g. the red and blue curves in the left panel of fig.~\ref{fig:GW-signal}). For comparison, we also show the GW signal for $\beta/H = 10, 30, 120$ in the right panel of fig.~\ref{fig:GW-signal} (corresponding to $N_c = 5$, shown by ``o'' in fig.~\ref{fig:beta-vs-TnTc}). For both choices of $N_c$ we see that while the signal strength is reduced and peak frequency is increased due to a non-zero $\epsilon_3$, there is still a possibility of discovery.

\section{General comments}
\label{sec:GeneralComments}

In this work we have argued for including self-interaction terms in the bulk potential of the stabilizing scalar. We considered a bulk potential with a quadratic and a cubic term, with signs chosen such that both terms grow in the IR. We stayed in the limit of small back-reaction and within the radion EFT. In the presence of a cubic term, and for a large hierarchy, the defining features of the radion potential are changed, and the radion mass is increased. The same effect also reduced the bounce action and thereby increased the rate for a transition from the hot phase to the RS phase. Equivalently, this reduced the amount of supercooling needed before completing the phase transition and increased the maximum number of colors $N_\text{max}$ for which the phase transition can complete, for a given amount of supercooling. For a choice of parameters, we were able to have $N_\text{max} \sim 10$. We also discussed the resulting GW signals from bubble collisions, and showed that in the presence of self-interactions, the parameter $\beta/H$ that characterizes the frequency and abundance of the GWs does not get too small, unlike the case when the CFT breaking in the IR is small and $\beta/H$ is close to zero.

\begin{figure}[ht!]
    \centering
    \includegraphics[width=0.95\textwidth]{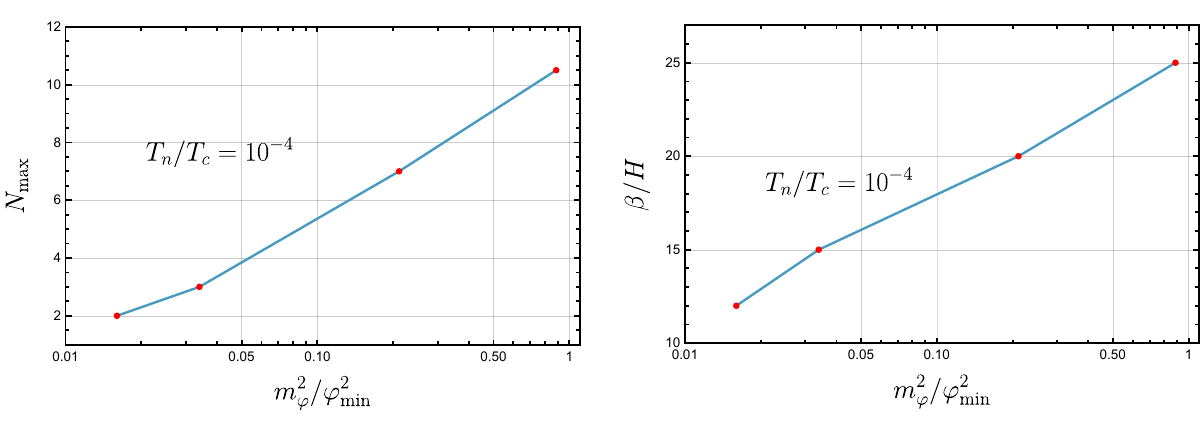}
    \caption{\small{The effect on $N_\text{max}$ (left) and $\beta/H$ (right) as the radion mass squared increases due to a non-zero $\epsilon_3$. The curves are shown for a fixed $T_n/T_c = 10^{-4}$.}}
    \label{fig:summary}
\end{figure}

In fig.~\ref{fig:summary} we show $N_\text{max}$ and $\beta/H$ for a moderate amount of supercooling $T_n/T_c = 10^{-4}$, as a function of the mass squared of the physical radion. The figure summarizes the main point of the paper---the presence of a self-interaction term increases the mass of the physical radion and the same effect also reduces the bounce action to increase $N_\text{max}$, while disfavoring a small $\beta/H$. As the radion mass gets close to the KK scale, we cannot trust the result entirely, and a complete 5D calculation would be necessary. Our conclusion is similar to~\cite{Csaki:2023pwy}, even though the underlying dynamics driving the radion mass up is different.  

Our work is a first step towards systematically including the effects of strong coupling in the IR, in the dynamics of the phase transition. In this work we have chosen parameters such that the radion is lighter than the other KK masses, allowing us to use the radion EFT. A more generic situation is when the radion mass is of the same order as other KK masses, which would drive the calculation out of the radion EFT and a full 5D calculation would be needed. The back-reaction would be important in the IR and it would change the free energy of both the RS and the BB phase. The stability of the phases can change the order of the phase transition, and a full 5D gravitational instanton computation would be needed to address the question properly. Rather than modeling choices, these are general effects. We will address some of these issues in a more general model in a future work.

\section*{Acknowledgements}
We would like to thank P. Creminelli, P. Du and A. Pomarol for comments and discussions at various stages. We would also like to thank N. DePorzio, J. Lodman and W. L. Xu for collaborations at an early stage. The work of RKM and LR is supported by the National Science Foundation under Grant Nos.~PHY-1620806, PHY-1748958 and PHY-1915071, the Chau Foundation HS Chau postdoc award, the Kavli Foundation grant “Kavli Dream Team,” and the Moore Foundation Award 8342. Part of this work was completed at Aspen Center for Physics, which is supported by National Science Foundation grant PHY-2210452.

\appendix

\section{Numerical and approximately analytical method}
\label{app:ApproxSols}
In this appendix we briefly outline the procedure to obtain the approximate analytical solutions for the scalar $\chi$, for both BB and RS backgrounds. We also describe the numerical method to obtain the profile for $\chi$ in the BB background.

The equation of motion for $\chi$ in the background of eq.~\eqref{eq:RSandAdSSmetric} is
\begin{align}
    \left(1-e^{4(r-r_h)}\right)\frac{\dd ^2 \chi}{\dd r^2} - 4 \frac{\dd \chi}{\dd r} - \frac{\dd V_B(\chi)}{\dd \chi} - \sum_i\delta(r-r_i)\frac{\dd V_i(\chi)}{\dd \chi} = 0\:,
\end{align}
where $i$ runs over $(uv, ir)$ for the RS case and only over $(uv)$ for the BB case. The range of $r$ is $0\leq r \leq r_\ir$ for the RS case and $0\leq r \leq r_h$ for the BB case. We will work with a rescaled coordinate $y=r/r_\ir$ for the RS case, and $y=r/r_h$ for the BB case, so that $0\leq y \leq 1$ for both backgrounds. Given the bulk and brane potentials in eq.~\eqref{eq:GW-bulkPotential}, \eqref{eq:GW-branePotential}, the equation and the boundary conditions for the RS and BB backgrounds are
\begin{align}
    \text{RS:}\qquad
    &\frac{1}{r_\ir^2}\frac{\dd ^2 \chi}{\dd y^2} - \frac{4}{r_\ir} \frac{\dd \chi}{\dd y} - 4\epsilon_2\chi - 4\epsilon_3\chi^2 = 0,\:\: \chi(0) = v_\uv,\:\:\chi'(1) = -r_\ir \alphair\:,
    \label{eq:GW-EOM-RS}
    \\
    \text{BB:}\qquad
    &\left(\frac{1-e^{4r_h(y-1)}}{r_h^2}\right)\frac{\dd ^2 \chi}{\dd y^2} - \frac{4}{r_h} \frac{\dd \chi}{\dd y} - 4\epsilon_2\chi - 4\epsilon_3\chi^2 = 0,\:\: \chi(0) = v_\uv\:.
    \label{eq:GW-EOM-BB}
\end{align}

\subsection{RS background}
In the limit $r_\ir\gg1, |\epsilon_2| r_\ir \lesssim 1, |\epsilon_3| r_\ir \lesssim 1$, eq~\eqref{eq:GW-EOM-RS} can be solved approximately by singular perturbation theory and boundary layer analysis~\cite{bender1999advanced}. Close to $r = y = 0$, $\chi$ varies slowly and the derivatives are small. To leading order in $1/r_\ir$, one can drop the second derivative in eq.~\eqref{eq:GW-EOM-RS}, which results in a first-order differential equation, readily solved. Applying the boundary condition $\chi(0) = v_\uv$ to the first-order differential equation resulting from dropping the second derivative term in eq.~\eqref{eq:GW-EOM-RS}, we get
\begin{align}
    \chi_\text{left}(y) = \frac{v_\uv e^{-\epsilon_2 r_\ir y}}{1 + v_\uv \epsilon_3 \left(\frac{1-e^{-\epsilon_2 r_\ir y}}{\epsilon_2}\right)}\:,
    \label{eq:GW-Sol-RS-LeftSide}
\end{align}
where the subscript ``left'' refers to the fact that the solution is valid only on the left side of the interval. As $y\to1$, $\chi$ has to change fast to match the boundary condition at $y=1$. In a small region near $y=1$, of width $\mathcal{O}(1/r_\ir)$, $\chi$ itself is small, but changes fast so that the derivatives are large. In this region (referred to as ``right'') one only keeps the first and the second derivative terms in eq.~\eqref{eq:GW-EOM-RS}, which gives
\begin{align}
    \chi_\text{right}(y) = -\frac{\alphair}{4}e^{4 r_\ir (y-1)} + C\:,
\end{align}
where we have applied the boundary condition $\chi'(1) = -r_\ir \alphair$, and $C$ is an undetermined constant at the moment. Using asymptotic matching in a region where both the solutions are valid, and requiring the same functional form, the constant $C$ can be fixed. The final solution is given as (now switching to $r= r_\ir\,y$)
\begin{align}
    \chi_\text{RS}(r) = -\frac{\alpha_\ir}{4} e^{4(r-r_\ir)} + \frac{v_\uv e^{-\epsilon_2 r}}{1 + v_\uv \epsilon_3 (\frac{1-e^{-\epsilon_2 r}}{\epsilon_2})}\:.
    \label{eq:GW-profile-RS1}
\end{align}
For consistency, we can check if the equations of motion and the boundary conditions are satisfied. In the limit of $r_\ir \gg 1, v_\uv \ll 1, |\epsilon_2|\ll1,|\epsilon_3|\ll1$, the errors on the boundary and in the bulk are small, and under control. Note that for $\epsilon_2 < 0$, the denominator of eq.~\eqref{eq:GW-Sol-RS-LeftSide} can vanish at some $r$. Clearly that is outside the validity of the approximation since the second derivative would not be small there anymore. To be consistent, we therefore need $-\epsilon_2 r_\ir < \log(1+\epsilon_2/v_\uv\epsilon_3)$. For $\epsilon_2<0, \epsilon_3<0$, in the limit of small $\epsilon_2$, this condition simplifies to $r_\ir < 1/v_\uv|\epsilon_3|$, which can be satisfied for small $v_\uv$. 

We can understand intuitively the role of $\epsilon_3$ as follows. For small enough $r$ one can expand the exponential in the denominator in the second term of eq.~\eqref{eq:GW-profile-RS1}, and approximate it as follows:
\begin{align}
    \frac{v_\uv e^{-\epsilon_2 r}}{1 + v_\uv \epsilon_3 (\frac{1-e^{-\epsilon_2 r}}{\epsilon_2})}
    \approx
    \frac{v_\uv e^{-\epsilon_2 r}}{1 + v_\uv \epsilon_3 r }
    \approx v_\uv e^{-(\epsilon_2+v_\uv\epsilon_3)r}\:.
\end{align}
This makes it clear that for small $r$, $\epsilon_3$ effectively increases $\epsilon_2$. This effect compounds as $r$ increases. 

\subsection{BB background}
Similar to the RS case, eq.~\eqref{eq:GW-EOM-BB} can be solved in the limit of large $r_h$. For $y = r/r_h \ll1$, the exponential term in the coefficient of $\chi''$ can be dropped, and we have the same solution as eq.~\eqref{eq:GW-Sol-RS-LeftSide}. Note that the UV boundary condition is the same for both RS and BB cases. As $y\to1$, \textit{unlike} the RS case, the coefficient of $\chi''$ becomes small, so that one can again drop the second derivative term. Since the only condition on $\chi(r)$ is to be regular at $y=1$, the leading order solution is
\begin{align}
    \chi_\text{BB}(r) = \frac{v_\uv e^{-\epsilon_2 r}}{1 + v_\uv \epsilon_3 (\frac{1-e^{-\epsilon_2 r}}{\epsilon_2})}\:.
    \label{eq:GW-profile-BB-approx}
\end{align}
To be more precise, we can do a Taylor series expansion around $y=1$ to solve the equation near $y=1$ and match it to $\chi_\text{left}(r)$ at some intermediate value. This procedure gives the same result as above, to leading order.

We would like to check the approximate solution~\eqref{eq:GW-profile-BB-approx} with the numerical solution for the profile. Without an explicit boundary condition to be applied at $y=1$, it is not clear a priori how to numerically solve for $\chi$ in the BB geometry. For this we apply a method that is akin to matching in an intermediate region. We solve the full equation numerically with given $\chi(0)$, and for some value of $\chi'(0)$, varying $\chi'(0)$. We also expand the equation in a Taylor series expansion around $y=1$ which analytically fixes $\chi'(1)$ in terms of $\chi(1)$. As we vary $\chi'(0)$ in the numerical solution, we check whether the analytical relation between $\chi(1)$ and $\chi'(1)$ is satisfied, which uniquely determines $\chi'(0)$, and hence the numerically regular solution.

In the special case of $\epsilon_3=0$, the solution is a linear combination of hypergeometric functions~\cite{Creminelli:2001th}, and the correct linear combination regular at the horizon is easily identified. Figure~\ref{fig:GW-profile-SI} shows a comparison between the solutions: for $\epsilon_3 \neq 0$, between the approximate solutions obtained in this appendix and the numerical solution obtained by the method discussed in the previous paragraph, and for $\epsilon_3 = 0$, between the exact solution and the approximate solution.

\section{Radion potential}
\label{app:RadionPotential}
Starting with the purely gravitational 5D action
\begin{align}
S = \int d^5x \sqrt{g}\left(-2M_5^3 \mathcal{R}[g] - \Lambda_5 \right) 
-\sum_{i=uv, ir}\int d^4x \sqrt{g_i} \: T_i\:,
\end{align}
where $-\Lambda_5=T_\uv = -T_\ir = 24 M_5^3$ (setting $\ell_\text{AdS} = 1$), and plugging back the metric
\begin{align}
ds^2 = -e^{-2r}dx^2 + dr^2 \: , \qquad 0 \leq r \leq r_\ir\:,
\end{align}
but making $r_\ir$ a 4D field $r_\ir(x)$, one generates a kinetic term for $r_\ir(x)$. In terms of $\varphi=\exp(-r_\ir)$, the 4D action looks like
\begin{align}
S = 
-12 M_5^3\int d^4 x (\partial \varphi)^2 \:.
\end{align}
At this point, there is no potential for $\varphi$ and it is a modulus. To generate a potential to stabilize the geometry, we add a GW scalar $\chi$ with the action defined in eq.~\eqref{eq:GW-action} and solve for the background value of $\chi$ which is a function of $r$ due to the choice of boundary and bulk potentials. Evaluating $S_\chi$ on $\chi(r)$ gives the potential for $\varphi$, for which we outline the steps now.  

Plugging $\chi(r)$ in $S_\chi$, we get
\begin{align}
    S_\chi &= \int \dd^4 x \int_0^{r_\ir} \dd r\, e^{-4r}\left(-\frac12\left(\frac{\dd\chi}{\dd r}\right)^2 - V_B(\chi)\right) - \sum_{i = uv, ir}e^{-4r_i}V_i(\chi(r_i)) \:.
    \label{eq:action-GW-eval-on-sol}
\end{align}
Using $\chi$ equations of motion $\chi''-4\chi'-\partial_\chi V_B(\chi)=0$ and an integration by parts, the bulk term of $S_\chi$ evaluates to
\begin{align}
    \int \dd^4 x
    \left(-\frac12\left.\left(e^{-4r}\chi\frac{\dd\chi}{\dd r}\right)\right|_{0}^{r_\ir} - \int_0^{r_\ir} \dd r\,
    e^{-4r}\left(V_B(\chi) -\frac12\chi\frac{\dd V_B}{\dd \chi}
    \right)
    \right)\:.
    \label{eq:action-GW-bulk-eval-on-sol}
\end{align}
Adding the contribution from the $uv, ir$ localized terms (in eq.~\eqref{eq:action-GW-eval-on-sol}) to the bulk contribution in eq.~\eqref{eq:action-GW-bulk-eval-on-sol}, the potential is given as
\begin{align}
    V(r_\ir) &= \frac12\left.\left(e^{-4r}\chi\frac{\dd\chi}{\dd r}\right)\right|_{0}^{r_\ir}   + \int_0^{r_\ir} \dd r\,
    e^{-4r}\left(V_B(\chi) -\frac12\chi\frac{\dd V_B}{\dd \chi}
    \right) + V_\uv(\chi(0)) + e^{-4r_\ir}V_\ir(\chi(r_\ir))\:.
    \label{eq:RadPotFull}
\end{align}
For Dirichlet boundary condition in UV, $V_\uv(\chi) = 0$. Given the bulk potential in eq.~\eqref{eq:GW-bulkPotential}, $V_B - (1/2) \chi \partial_\chi V_B = -(2/3) \epsilon_3 \chi^3$. Using $\chi(r)$ from eq.~\eqref{eq:GW-profile-RS}, we need to evaluate the integral
\begin{align}
    -\frac23\epsilon_3\,\int_0^{r_\ir}\dd r\, e^{-4r}
    \left(
    -\frac{\alpha_\ir}{4} e^{4(r-r_\ir)} + \frac{v_\uv e^{-\epsilon_2 r}}{1 + v_\uv \epsilon_3 \left(\frac{1-e^{-\epsilon_2 r}}{\epsilon_2}\right)}
    \right)^3\:.
    \label{eq:GW-Pot-bulk}
\end{align}
To proceed, we need some identities related to hypergeometric functions. First note that each term in the above is an integral of a general form, with a closed-form answer expressed in terms of hypergeometrics:
\begin{align}
    \int \dd r \frac{e^{ar}}{(1+b e^{cr})^n} = \frac{e^{ar}}{(1+b e^{cr})^{n-1}}\,{}_2F_1(1,\, 1 - n + \frac{a}{c},\, 1 + \frac{a}{c},\, be^{cr})\: .
    \label{eq:2F1Integral}
\end{align}
Two other useful identities are~\cite{Cvitkovi__2017}:
\begin{align}
    {}_2F_1(a,b,c;z) &= (1-z)^{c-a-b}{}_2F_1(c-a, c- b, c; z)\:,
    \nonumber \\
    {}_2F_1(a+\delta\lambda,b,c+\lambda)&\approx \left(1-\delta z\right)^{-b}\:,\qquad
    |\lambda|\gg1\:,|\delta|\leq 1\:.
    \label{eq:2F1Identities}
\end{align}
To simplify the further expressions, we define
\begin{align}
    \epsilon_{32} = \frac{\epsilon_3}{\epsilon_2}\:,\:\: \lambda = \frac{v_\uv\epsilon_{32}}{1+v_\uv\epsilon_{32}}\:,\:\:
    Y(r) = \frac{\lambda e^{-\epsilon_2 r}}{1-\lambda e^{-\epsilon_2 r}} 
    \:,
    \label{eq:RadPotCalcNotation}
\end{align}
in terms of which $\chi(r)$ can be written as
\begin{align}
    \chi(r) = -\frac{\alphair}{4}e^{4(r-r_\ir)}+\frac{1}{\epsilon_{32}}Y(r)\:.
\end{align}
We now expand the integrand in eq.~\eqref{eq:GW-Pot-bulk}, use eqs.~\eqref{eq:2F1Integral},~\eqref{eq:2F1Identities} for simplification and the notation of eq.~\eqref{eq:RadPotCalcNotation}. Calling the four terms from expanding eq.~\eqref{eq:GW-Pot-bulk} as $t_1, t_2, t_3, t_4$, we have 
\begin{align}
    t_1 &= -\frac23\epsilon_3\,\int_0^{r_\ir}\dd r\, e^{-4r}
    \left(
    -\frac{\alpha_\ir}{4} e^{4(r-r_\ir)}\right)^3
    =\frac{1}{96}\epsilon_3\alphair^3 e^{-12r_\ir}\int_0^{r_\ir} \dd r e^{8r}
    \nonumber \\
    &=\frac{1}{768}\epsilon_3\alphair^3\left(\varphi^4-\varphi^{12}\right)\:.
    \\
    t_2 &= -\frac23\epsilon_3\,\int_0^{r_\ir}\dd r\, e^{-4r}
    \,3\,\left(-\frac{\alpha_\ir}{4} e^{4(r-r_\ir)}\right)
    \left(\frac{1}{\epsilon_{32}}Y(r)\right)^2
    \nonumber \\
    &=\frac12\frac{\epsilon_2^2}{\epsilon_3}\alphair\,e^{-4r_\ir}\,
    \int_0^{r_\ir}\dd r \frac{\lambda^2e^{-2\epsilon_2 r}}{(1-\lambda e^{-\epsilon_2 r})^2}
    =\frac12\frac{\epsilon_2}{\epsilon_3}\alphair\,e^{-4r_\ir}\,
    \left.\Big(-Y(r)+\log(1+Y(r))\Big)\right|_{0}^{\rir}
    \nonumber \\
    &=\frac12\frac{\epsilon_2}{\epsilon_3}\alphair\varphi^4
    \left(-Y(\rir)+\log(1+Y(\rir))+\frac{\lambda}{1-\lambda}+\log(1-\lambda)\right)\:.
    \\
    t_3 &=-\frac23\epsilon_3\,\int_0^{r_\ir}\dd r\, e^{-4r}
    \,3\,\left(-\frac{\alpha_\ir}{4} e^{4(r-r_\ir)}\right)^2
    \left(\frac{1}{\epsilon_{32}}Y(r)\right)
    \nonumber \\
    &=-\frac18\epsilon_2\alphair^2e^{-8\rir}
    \int_0^{r_\ir}\dd r \frac{\lambda e^{(4-\epsilon_2) r}}{(1-\lambda e^{-\epsilon_2 r})}
    =\frac{1}{32}\epsilon_2\alphair^2e^{-8\rir}
    \left.\left(e^{4r}{}_2F_1(1, \frac{4}{\epsilon_2},1+\frac{4}{\epsilon_2};\lambda^{-1}e^{\epsilon_2 r})\right)\right|_{0}^{r_\ir}
    \nonumber \\
    &= \frac{1}{32}\epsilon_2\alphair^2e^{-8\rir}
    \left.\Big(-e^{4r}Y(r)\Big)\right|_{0}^{r_\ir}
    =\frac{1}{32}\epsilon_2\alphair^2
    \left(-\varphi^4Y(\rir)+\varphi^8\left(\frac{\lambda}{1-\lambda}\right)\right)\:.
    \\
    t_4 &=-\frac23\epsilon_3\,\int_0^{r_\ir}\dd r\, e^{-4r}
    \left(\frac{1}{\epsilon_{32}}Y(r)\right)^3
    =-\frac23\frac{\epsilon_2^3}{\epsilon_3^2}
    \int_0^{r_\ir}\dd r \frac{\lambda^3 e^{-(4+3\epsilon_2) r}}{(1-\lambda e^{-\epsilon_2 r})^3}
    \nonumber\\
    &=-\frac{1}{6}\frac{\epsilon_2^3}{\epsilon_3^2}
    \left.\left(e^{-4r}{}_2F_1(3, -\frac{4}{\epsilon_2},1-\frac{4}{\epsilon_2};\lambda^{-1}e^{\epsilon_2 r})\right)\right|_{0}^{r_\ir}
    =-\frac{1}{6}\frac{\epsilon_2^3}{\epsilon_3^2}
    \left.\Big(-e^{-4r}Y^3(r)\Big)\right|_{0}^{r_\ir}
    \nonumber \\
    &=\frac16\frac{\epsilon_2^3}{\epsilon_3^2}
    \left(\varphi^4Y^3(\rir)-\left(\frac{\lambda}{1-\lambda}\right)^3\right)\:.
\end{align}
We have used $|\epsilon_2|\ll1$ in the above to simplify the hypergeometric functions. Collecting everything together, keeping to leading order in $\varphi$, and dropping overall constants, the integral in eq.~\eqref{eq:GW-Pot-bulk} is given as:
\begin{align}
    &\varphi^4\left(A+BY+CY^3+D\log(1+Y)\right)\:,\:\: Y = \frac{\lambda \varphi^{\epsilon_2}}{1-\lambda \varphi^{\epsilon_2}}
    \nonumber \\
    &A=\frac{1}{768}\epsilon_3\alphair^2 -\frac{\lambda}{1-\lambda}+\log(1-\lambda)\:,
    \:\:B = -\frac{1}{32}\epsilon_2\alphair^2-\frac12\frac{\epsilon_2}{\epsilon_3}\alphair\:,
    \:\:C = \frac16\frac{\epsilon_2^3}{\epsilon_3^2}\:,
    \:\:D = \frac12\frac{\epsilon_2}{\epsilon_3}\alphair\:.
\end{align}
In addition to all the terms in eq.~\eqref{eq:RadPotFull}, there is a potential generated by the detuning of the IR brane tension, and is given by
\begin{align}
    V_\text{detune}(\varphi) &= \widetilde{\tau}\,\varphi^4\:,
\end{align}
where $\widetilde{\tau}$ is a free parameter at this point. Together with above, including all the contributions in eq.~\eqref{eq:RadPotFull}, and keeping to linear order in $\epsilon_2$, the radion potential is given as
\begin{align}
    &V(\varphi) =\varphi^4\left(a_1+a_2Y+a_3\log(1+Y)\right)\:,\:\: Y = \frac{\lambda \varphi^{\epsilon_2}}{1-\lambda \varphi^{\epsilon_2}}
    \nonumber \\
    &a_1 = \tau\:,\:\: 
    a_2= -\frac{1}{32}\epsilon_2\alphair^2-\frac{\epsilon_2}{\epsilon_3}\alphair + 2\alphair\:,\:\:
    a_3= \frac12\frac{\epsilon_2}{\epsilon_3}\alphair\:,
\end{align}
where we have absorbed terms to define an overall $\tau$ and used exact values for $\chi(0)$ and $\chi'(r_\ir)$ (these are the specified boundary conditions and are only approximately satisfied by the approximate solution). The $\epsilon_3 \to 0$ limit is finite, and in this limit $a_2Y+a_3\log(1+Y)$ reduces to $-(1/2)\alphair v_\uv \varphi^{\epsilon_2}$. Note that for $Y$ to stay finite, we need $r_\ir < (1/\epsilon_2)\log\lambda$.

To make it clear what are the reasonable values for parameters, and to keep the $N$ dependence of the dual theory clear, we define $\tau \equiv 24 M_5^3 \kappa^4$. Since $M_5^3 \sim \frac{N^2}{16\pi^2}$, this makes it clear that $\tau \sim \frac{N^2}{16\pi^2} \kappa^4$ (which is the glueball normalization) and we need $\kappa \lesssim 1$ for small back-reaction. The radion potential can be rewritten as
\begin{align}
    V(\varphi) &= 24 M_5^3\, \kappa^4\, \varphi^4\, v(\varphi/\varphi_\text{min}; \epsilon_2, \epsilon_3, \alphair, \vuv)\:,
\end{align}
where the function $v$ does not have an analytical expression (since $\varphi_\text{min}$ does not have an analytical expression), but is easily obtained numerically. The dimensionless function $v$ encodes the information about the breaking of the CFT, and plays a crucial role when calculating the bounce action. Putting everything together, the radion action is
\begin{align}
S = 
24 M_5^3\int \dd^4 x \left(-\frac12(\partial \varphi)^2 - \kappa^4 \,\varphi^4 \,v(\varphi/\varphi_\text{min}; \epsilon_2, \epsilon_3, \alphair, \vuv)\right) \:.
\end{align}

\section{Bounce action in the thin and thick wall limits}
\label{app:BounceAction}
Starting with the action 
\begin{align}
    S &= \int \dd^4 x \left(-12 M_5^2(\partial\varphi)^2 - V(\varphi)\right)\:,
\end{align}
the bounce action is obtained by looking for a solution to $\varphi$ that minimizes the Euclidean action
\begin{align}
    S_b & = \int \dd^4 x \left(12 M_5^2(\partial\varphi)^2 + \widetilde{V}(\varphi)\right)\:,\qquad
    \widetilde{V}(\varphi) = V(\varphi)- C 
    \:,
\end{align}
and evaluating the action on the solution. The constant $C$ is chosen to subtract the contribution from the false vacuum, and is given by $-2\pi^4 M_5^3 T_c^4 = V(\varphi_\text{min})$ for the thin wall case, and $-2\pi^4 M_5^3 T^4$ for the thick wall case. At zero temperature, $\varphi$ is assumed to be a function of the combination $\rho^2 = \vec{x}\cdot\vec{x} + t_E^2$ and this is referred as the $O(4)$ symmetric solution. At finite temperature, there is another saddle that can dominate. For inverse temperature $\beta = 1/T$, the Euclidean time is made periodic with period $\beta$. The field $\varphi$ is assumed to be a function of the combination $r^2 = \vec{x}\cdot\vec{x}$ and this is referred to as the $O(3)$ solution. In rest of the discussion we focus on the $O(3)$ solution only.

For $O(3)$ symmetric solutions, the action is given more explicitly as (ignoring temperature corrections to the potential)
\begin{align}
     S_b & = \int_0^\beta \dd t_E \int \dd^3 x \left(12 M_5^2(\partial\varphi)^2 +\widetilde{V}(\varphi)\right) = \frac{4\pi}{T}\int\dd r\, r^2 \left(12 M_5^2\left(\frac{\dd \varphi}{\dd r}\right)^2 +\widetilde{V}(\varphi)\right)\:.
     \label{eq:BounceAction-O3}
\end{align}
The equations of motion are
\begin{align}
    24 M_5^3 \left(\frac{\dd^2\varphi}{\dd r^2} + \frac{2}{r}\frac{\dd\varphi}{\dd r}\right) = \frac{\dd \widetilde{V}(\varphi, T)}{\dd \varphi} \approx \frac{\dd \widetilde{V}(\varphi)}{\dd \varphi}\:.
    \label{eq:EOM-full-bounceAction}
\end{align}
One of the boundary conditions is $\varphi' (r=0) = 0$ and we will have more to say about the second boundary condition later. 
\subsection{Thin wall case}
In the limit of the two vacua being very degenerate, the solution for $\varphi$ is such that it has a large region where it is constant (and equal to its value in the true vacuum) and then changes quickly to the value in the false vacuum. In this limit, the wall is small compared to the size of the bubble, hence this is called the thin-wall solution. In this limit, one can ignore the first derivative term in the equations of motion. Using the identity $2\dd^2y/\dd x^2 = \dd/\dd y (\dd y/\dd x)^2$, the equations can be solved to give
\begin{align}
    12 M_5^3\left(\frac{\dd\varphi}{\dd r}\right)^2 = \widetilde{V}(\varphi)\:,
    \label{eq:thin-wall-EOM}
\end{align}
where we used the boundary condition $\varphi'(r=0) = 0$ and the fact that at $r=0$, $\varphi = \varphi_\text{min}$ and $\widetilde{V}(\varphi_\text{min}) = 0$. To evaluate the action on this solution, a somewhat indirect approach is more intuitive. First note that for a bubble of size $R$, the field is mostly constant for $r\lesssim R$, changes quickly in the vicinity of $r=R$, and is again constant afterwards. We can split the action into these three regions. In the region $r\lesssim R$, we can drop the derivative, $\widetilde{V}(\varphi)$ is a constant, and the integrand is proportional to $r^2$. For $r\sim R$, the factor of $r^2$ can be approximated to be $R^2$, and we have to keep both the derivatives and the potential inside the integral. For $r\gtrsim R$ there is no contribution. We therefore have
\begin{align}
    S_b = \frac{4\pi}{T}\left(\widetilde{V}(\varphi)\int_0^R \dd r \, r^2 + R^2\int_{r\sim R} \dd r\left(12 M_5^3\left(\frac{\dd \varphi}{\dd r}\right)^2 + \widetilde{V}(\varphi)\right)\right)\:.
    \label{eq:Sb-as-a-function-of-R}
\end{align}
Using $\widetilde{V}(\varphi) = \Delta F = F_\text{FV} - F_\text{TV} > 0$ as the difference in the free energies between the two vacua, we get
\begin{align}
    S_b = \frac{4\pi}{T}\left(\frac13 \Delta F R^3 + R^2 S_1 \right)\:,\:\:\: 
    S_1 = \int_{r\sim R} dr \left(12 M_5^3\left(\frac{\dd \varphi}{\dd r}\right)^2 + \widetilde{V}(\varphi)\right)\:.
    \label{eq:Sb-and-S1}
\end{align}
Changing variables from $r$ to $\varphi$, $S_1$ can be written as
\begin{align}
    S_1 &= \sqrt{48M_5^3}\int_{0}^{\varphi_\text{min}} \dd \varphi \sqrt{\widetilde{V}(\varphi)}\:,
    \label{eq:S1}
\end{align}
and is independent of $R$. Since $\varphi$ changes from $0$ to $\varphi_\text{min}$ in the $r\sim R$ region, this fixes the limits of the integration in eq.~\eqref{eq:S1}.\footnote{Technically the lower limit is $\varphi\sim T$, since we are estimating the action from region $iii)$ (see discussion in main text). Taking the lower limit to be approximately zero does not change the estimate.} Equation~\eqref{eq:Sb-and-S1} gives $S_b$ as a function of $R$, which is minimized at $R = -2S_1/\Delta F$, at which the bounce action is given as
\begin{align}
    S_b = \frac{16\pi}{3T}\frac{S_1^3}{(\Delta F)^2}\:.
\end{align}
The free energy difference $\Delta F$ can be written in terms of the critical temperature $T_c$ as
\begin{align}
    \Delta F &= F_\text{FV} - F_\text{TV} = 2 \pi^4 M_5^3 \left(T^4- T_c^4\right)\:,
\end{align}
using which, the bounce action becomes
\begin{align}
    S_b &= \frac{4 }{3 \pi^7M_5^6} 
    \left(\frac{S_1}{T_c^3}\right)^3
    \frac{T_c/T}{\left(1-T^4/T_c^4\right)^2}\:.
\end{align}
One can also calculate the $\varphi$ profile for thin-wall case, using eq.~\eqref{eq:thin-wall-EOM} as
\begin{align}
    -\sqrt{12 M_5^3}\int_{\varphi_\text{min}}^\varphi \frac{\dd \varphi}{\sqrt{\widetilde{V}(\varphi)}} = r\:,
\end{align}
where we have chosen the negative sign of the square root, and used the boundary condition $\dd\varphi/\dd r=0$ at $r=0$, which by eq.~\eqref{eq:thin-wall-EOM} is at $\varphi = \varphi_\text{min}$.
\subsection{Thick wall case}
In the thick wall case, we minimize the action
\begin{align}
    S_b & = \int \dd^4 x \left(12 M_5^2(\partial\varphi)^2 + V(\varphi)+2\pi^4M_5^3T^4\right)\:,
\end{align}
One of the boundary conditions to be satisfied is the standard one: $\varphi'(0) = 0$ ($r$ being a radial coordinate). The second boundary condition is more subtle. The usual second boundary condition is to require $\varphi(r\to \infty) = \varphi_\text{false vacuum}$. In the case at hand, $\varphi$ is not a dynamical variable in the other phase. Inside the bubble, close to the boundary, we have $\varphi \sim T \approx 0$, and the energy is only in the gradient. Outside the bubble and close to the boundary, the energy is proportional to $T^4$. Requiring the energies to match at the bubble boundary we get~\cite{Agashe:2019lhy}
\begin{align}
    12 M_5^3 \left.\left(\frac{\dd \varphi}{\dd r}\right)^2\right|_{\varphi \, \approx \, 0} &= 2 \pi^4 M_5^3 T^4 \:.
\end{align}
We choose the negative sign in the square root because $\varphi$ starts near $\varphi_\text{min}$ at the center of the bubble and decreases to zero at the boundary, thereby the derivative is negative when $\varphi\to0$. Putting everything together, the equations and the boundary conditions are
\begin{align}
    &24 M_5^3 \left(
    \frac{\dd^2\varphi}{\dd r^2}
    +
    \frac{2}{r}\frac{\dd\varphi}{\dd r}
    \right)
    =
    \frac{\dd \widetilde{V}(\varphi)}{\dd \varphi}\:,\qquad
    \left.\frac{\dd \varphi}{\dd r}\right|_{r=0} = 0,\:\:
    \left.\frac{\dd \varphi}{\dd r}\right|_{\varphi\approx0} = -\frac{\pi^2}{\sqrt{6}}T^2\:.
\end{align}
Since the location at which $\varphi \approx 0$ is not known a priori, one has to approach the problem indirectly. The strategy is to solve the differential equation numerically with $\varphi(0) = \varphi_0, \varphi'(0) = 0$, calculate $r_*$ such that $\varphi(r_*)\approx0$, and adjust $\varphi_0$ till $\varphi'(r_*)$ has the appropriate value. For numerical convenience we define
\begin{align}
    &\frac{1}{24M_5^3}\widetilde{V}(\varphi) \equiv \kappa^4\, \varphi^4\, v(\varphi)\:,
    \nonumber \\
    &y = \kappa\, \varphi\, T^{-1}\:,\:\: x = \kappa\, r \, T\:.
\end{align}
With this rescaling, the action in eq.~\eqref{eq:BounceAction-O3} looks like
\begin{align}
    S_b & = \frac{96 \pi M_5^3}{\kappa^3} \int \dd x \, x^2 \left(\frac12\left(\frac{\dd y}{\dd x}\right)^2 
    +
    y^4 \, v(T y/\kappa)
    +
    \frac{\pi^4}{12}
    \right)\:.
\end{align}
The differential equation and the boundary conditions in the rescaled coordinates are
\begin{align}
    \frac{\dd^2y}{\dd x^2} + \frac{2}{x}\frac{\dd y}{\dd x} = \frac{\dd }{\dd y} y^4 v(Ty/\kappa)\:,
    \:\: 
    \left.
    \frac{\dd y}{\dd x}
    \right|_{x = 0} = 0\:,
    \:\: 
    \left.
    \frac{\dd y}{\dd x}
    \right|_{y \approx 0} = -\frac{\pi^2}{\sqrt{6}}\:.
    \label{eq:thick-wall-EOM}
\end{align}
As discussed before, since the value of $x$ at which $y\approx0$ is not known before solving the equation, we trade that boundary condition with $y(0) = y_0$, calculate $x_*$ such that $|y(x_*)| <\delta = 10^{-1}$ and adjust $y_0$ till $y'(x_*) + \pi^2/\sqrt{6} = 0$. Note that eq.~\eqref{eq:thick-wall-EOM} depends on $T/T_c$ and one has to solve it for different values of $T/T_c$ to get the temperature dependence of $S_b$. 

\begin{figure}[h]
    \centering
    \includegraphics[scale=0.5]{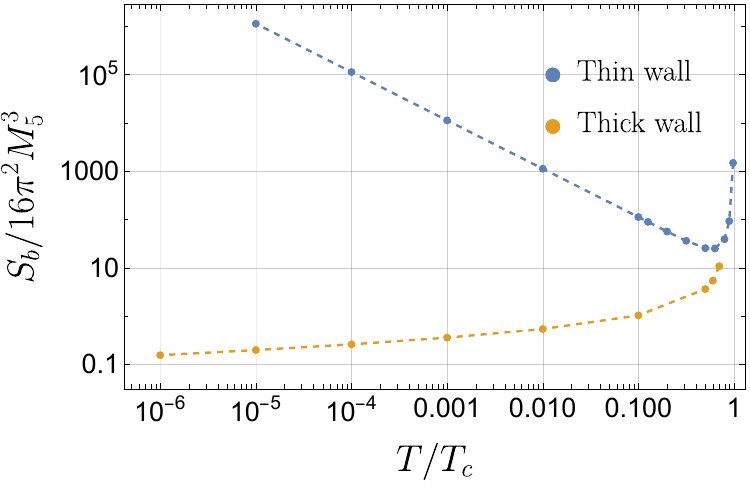}
    \:\:
    \includegraphics[scale=0.48]{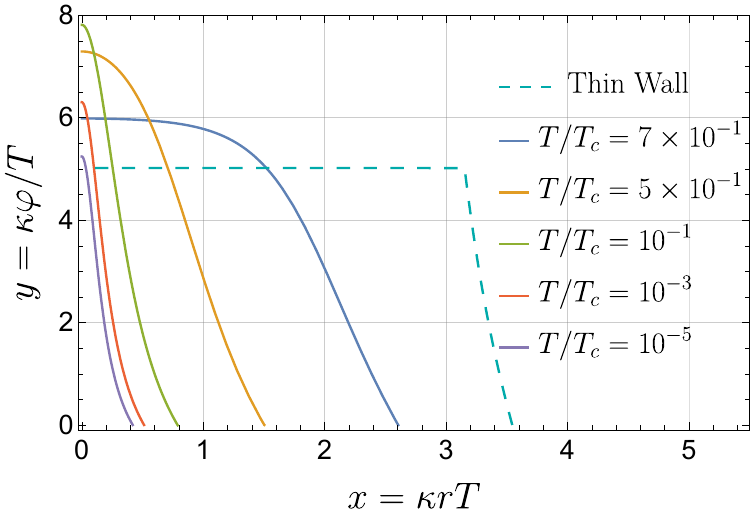}
    \caption{\small{Thin vs thick wall for $\tau = 2, \alphair = 5, \vuv = 1/5, \epsilon_2 = -1/10, \epsilon_3 = 0$. Left shows the bounce action and right shows the bubble profile.}}
    \label{fig:thin-vs-thick-Wall}
\end{figure}

Figure~\ref{fig:thin-vs-thick-Wall} compares the results for the thin and thick wall cases for a generic choice of parameters: left shows the bounce action as a function of $T/T_c$ for the thin and the thick wall cases, right shows the scalar profile for several values of $T/T_c$ for the thick wall case, and for the thin wall case where $T\approx T_c$ (the two vacua being almost degenerate). As the thin-wall limit is approached (i.e. the bubble radius gets bigger), the results for the thick-wall calculation converge to the thin-wall results.

When computing the bounce action and the bubble profiles for $\epsilon_3 \neq 0$, one technical difficulty is to be kept in mind. As discussed earlier, the radion potential $V(\varphi)$ has a singularity at a finite but non-zero $\varphi_s = (1/\lambda)^{1/\epsilon_2}$, where $\lambda = v_\uv (\epsilon_3/\epsilon_2)/(1+v_\uv(\epsilon_3/\epsilon_2))$. 
Given that $T_c \lesssim \varphi_\text{min}$, if the thick wall profile is computed at $T/T_c \ll 1$, the relevant $\varphi$ probed is of the order $(T/T_c)\varphi_\text{min} \ll \varphi_\text{min}$, and for small enough $T$, one can be sensitive to the singular point $\varphi = \varphi_s$. Since $\varphi_s \ll \varphi_\text{min}$, one can calculate for intermediate temperatures without running into this issue. In this work we consider only values of $T/T_c$ such that we are not sensitive to the singularity.
Further, note that the boundary condition requires imposing a condition when $y\approx0$ or equivalently $\varphi\approx0$. Numerically we require $\left|y(x_*)\right| \leq \delta = 10^{-1}$ or equivalently $\left|\varphi\right| \leq T (\delta/\kappa) \sim 10^{-1} T$. As long as $T$ is not too small, we do not probe the singular region of the potential.

\bibliographystyle{utphys}
\bibliography{references}
 
\end{document}